\documentclass[sigconf, nonacm]{acmart}

\newcommand\vldbdoi{XX.XX/XXX.XX}

\DeclareMathAlphabet{\altmathcal}{OMS}{zplm}{m}{n}
\usepackage{epsfig,endnotes,xspace,hyperref,url,diagbox}
\AtBeginDocument{}
\usepackage{amsmath, amsfonts}
\usepackage{dsfont}
\usepackage{tikz}
\usepackage{enumerate}
\usepackage{varioref}
\usepackage{graphicx}
\usepackage{epstopdf}
\usepackage{color}
\usepackage{multirow}
\usepackage{subfigure}
\usepackage{comment}
\usepackage[utf8]{inputenc}
\usepackage{mathtools}
\usepackage{wrapfig}
\usepackage{algorithm}
\usepackage[noend]{algpseudocode}

\usepackage{caption}

\usepackage{cryptocode}

\newcommand{\mypara}[1]{\vspace*{0.05in}\noindent\textbf{#1.}$\;$}

\definecolor{revision}{RGB}{0,0,255}

\newcommand{\revisionstart}{\begin{color}{revision}}
\newcommand{\revisionend}{~\!\!\end{color}}

\newtheorem{definition}{Definition}

\newtheorem{example}{Example}

\newcommand{\ind}[1]{\mathds{1}_{\{#1\}}}

\newcommand{\Var}[1]{\ensuremath{\mathsf{Var}\left[#1\right]}\xspace}

\renewcommand{\Pr}[1]{\ensuremath{\mathsf{Pr} \left[#1\right] }\xspace}
\newcommand{\tuple}[1]{\ensuremath{\langle #1\rangle}\xspace}

\renewcommand{\AA}{\mathbf{A}}

\newcommand{\results}{\ensuremath{\mathsf{\mathbf{R}}}\xspace}

\newcommand{\olh}{\ensuremath{\mbox{OLH}}\xspace}

\newcommand{\grr}{\ensuremath{\mbox{GRR}}\xspace}

\newcommand{\ep}[2][]{{\bf E}_{#1}\hspace{-0.06cm}\left[#2\right]\xspace}

\newcommand{\epinline}[1]{{\bf E}[#1]\xspace}

\newcommand{\mytitle}{High-Dimensional Range Query under Local Differential Privacy}
\renewcommand{\mytitle}{Answering Multi-Dimensional Range Queries under \\ Local Differential Privacy}

\usepackage{booktabs}
\newcommand{\figurepath}[1]{figure/#1}

\newcommand{\rv}[1]{\textcolor{black}{#1}}

\def\fwidth{0.48}
\def\fheight{0.27}

\def\fmpagewidth{0.48}

\def\crf{\Phi} 

\def\sds{\rv{D_{\eta}}}

\def\rf{\bar f}

\def\eef{f} 

\def\sums{Y} \title{\mytitle}

\begin{document}

\author{
	Jianyu Yang{$^{1,2*}$},
	Tianhao Wang{$^{2}$},
	Ninghui Li{$^{2}$},
	Xiang Cheng{$^{1}$},
	Sen Su{$^{1}$}
}
\affiliation{
{$^{1}$}\textit{{State Key Laboratory of Networking and Switching Technology}},\\
\textit{{Beijing University of Posts and Telecommunications, Beijing, China}}\\
\textit{{\{jyyang, chengxiang, susen\}@bupt.edu.cn}}\\
{$^{2}$}\textit{{Department of Computer Science, Purdue University, West Lafayette, USA}}\\
\textit{\{yang1896, tianhaowang\}@purdue.edu, ninghui@cs.purdue.edu}
}

\begin{abstract}

	In this paper, we tackle the problem of answering multi-dimensional range queries under local differential privacy. There are three key technical challenges: capturing the correlations among attributes, avoiding the curse of dimensionality, and dealing with the large domains of attributes. None of the existing approaches satisfactorily deals with all three challenges. Overcoming these three challenges, we first propose an approach called Two-Dimensional Grids (TDG). Its main idea is to carefully use binning to partition the two-dimensional (2-D) domains of all attribute pairs into 2-D grids that can answer all 2-D range queries and then estimate the answer of a higher dimensional range query from the answers of the associated 2-D range queries. However, in order to reduce errors due to noises, coarse granularities are needed for each attribute in 2-D grids, losing fine-grained distribution information for individual attributes. To correct this deficiency, we further propose Hybrid-Dimensional Grids (HDG), which also introduces 1-D grids to capture finer-grained information on distribution of each individual attribute and combines information from 1-D and 2-D grids to answer range queries. To make HDG consistently effective, we provide a guideline for properly choosing granularities of grids based on an analysis of how different sources of errors are impacted by these choices. Extensive experiments conducted on real and synthetic datasets show that HDG can give a significant improvement over the existing approaches.

\end{abstract}
\maketitle

\pagestyle{plain}

\renewcommand{\thefootnote}{\fnsymbol{footnote}}
\footnotetext[1]{Work done while studying as a visiting student at Purdue University.\\\noindent\rule[0.25\baselineskip]{\columnwidth}{0.15mm}}

\begingroup
\renewcommand\thefootnote{}\footnote{\noindent
	This work is XXXXXXXX XXXXXXXXXXXXXXXXXXXXXXXXXXXX. Visit \url{XXXXXXXXXXXXXXXXXXXXXXXXXXXXXXXXXXXXXXXX} to view XXXXXXXXXXXXXXX. XXXXXXXXXXXXXXXXXXXXXXXXXXXXXXXXX \href{mailto:XXXXXXXXXX}{XXXXXXXXX}. Copyright is held by the owner/author(s). XXXXXXXXXXXXXXXX XXXXXXXXXXXXXXXXXXXXXXXXX. \\
	\raggedright Proceedings of XXXXXXXXXXXXXX. \\
	\href{https://doi.org/\vldbdoi}{doi:\vldbdoi} \\
}\addtocounter{footnote}{-1}\endgroup

\section{Introduction}
\label{sec:intro}

Nowadays, users' data records contain many ordinal or numerical attributes in nature, e.g., income, age, the amount of time viewing a certain page, the number of times performing a certain actions, etc. The domains of these attributes consist of values that have a meaningful total order. A typical kind of fundamental analysis over users' records is multi-dimensional range query, which is a conjunction of multiple predicates for the attributes of interest and asks the fraction of users whose record satisfies all the predicates.
In particular, a predicate is a restriction on the range of values of an attribute.
However, users' records regarding these ordinal attributes are highly sensitive. Without strong privacy guarantee, answering multi-dimensional range queries over them will put individual privacy in jeopardy. Thus, developing effective approaches to address such privacy concerns becomes an urgent need.

In recent years, local differential privacy (LDP) has come to be the \emph{de facto} standard for individual privacy protection.
Under LDP, random noise is added on the client side before the data is sent to the central server.  Thus, users do not need to rely on the trustworthiness of the central  server.  This desirable feature of LDP has led to wide deployment by industry (e.g., by Google~\cite{DBLP:conf/ccs/ErlingssonPK14}, Apple~\cite{url:apple}, and Microsoft~\cite{DBLP:conf/nips/DingKY17}). However, existing LDP solutions~\cite{DBLP:journals/pvldb/CormodeKS19,DBLP:conf/sigmod/Li0LLS20,DBLP:conf/sigmod/WangDZHHLJ19} are mostly limited to one-dimensional (1-D) range queries on a single attribute and cannot be well extended to handle multi-dimensional range queries.

In this paper, we tackle the problem of answering multi-dimens-ional range queries under LDP. Given a large number of users who have a record including multiple ordinal attributes, an untrusted aggregator aims at answering all possible multi-dimensional range queries over the users' records while satisfying LDP. To address the problem, we identify three key technical challenges:
1) how to capture the correlations among attributes,
2) how to avoid the curse of dimensionality, and
3) how to cope with the large domains of attributes.
Any approach failing to solve any of these three challenges will have poor utility.
As we show in Section~\ref{sec:baseline_approaches}, none of the existing approaches or their extensions can deal with all three challenges at the same time.

Overcoming these three challenges, we first propose an approach called Two-Dimensional Grids (TDG).
Its main idea is to carefully use binning to partition the two-dimensional (2-D) domains of all attribute pairs into 2-D grids that can answer all possible 2-D range queries and then estimate the answer of a higher dimensional range query from the answers of the associated 2-D range queries.
\rv{However, in order to reduce errors due to noises, coarse granularities are needed for each attribute in 2-D grids, losing fine-grained distribution information for individual attributes.
	When computing the answer of a 2-D range query by the cells that are partially included in the query, it needs to assume a uniform distribution within these cells, which may lead to large errors.}
\rv{To correct this deficiency, we further propose an upgraded approach called Hybrid-Dimensional Grids (HDG), whose core idea is combining hybrid dimensional (1-D and 2-D) information for better estimation. In particular, HDG also introduces 1-D grids to capture finer-grained information on distribution of each individual attribute and combines information from 1-D and 2-D grids to answer range queries. In both TDG and HDG, users are divided into groups, where each group reports information for one grid. After collecting frequencies of cells in each grid under LDP, the aggregator uses techniques to remove negativity and inconsistency among grids, and finally employs these grids to answer range queries.}

However, it is still nontrivial to make HDG consistently effective, since the granularities for 1-D and 2-D grids can directly affect the performance of HDG. Consequently, it is essential to develop a method for determining the appropriate grid granularities so that HDG can guarantee the desirable utility. In particular in HDG, there are two main sources of errors: those due to noises generated by the random nature of LDP and those due to binning.
When the distribution of values is fixed, errors due to binning do not change and can be viewed as bias because of the uniformity assumption, and errors due to noises can be viewed as variance.  Thus choosing the granularities of grids can be viewed as a form of bias-variance trade-off.  Finer-grained grids lead to greater error due to noises, while coarser-grained ones result in greater error due to biases. The effect of each choice depends both on the privacy budget $\varepsilon$, population, and property of the distribution. By thoroughly analyzing the two sources of errors, we provide a guideline for properly choosing grid granularities under different parameter settings.

\rv{By capturing the necessary pair-wise attribute correlations via 2-D grids, both approaches overcome the first two challenges. Moreover, since they properly use binning with the provided guideline to reduce the error incurred by a large domain, the third challenge is carefully solved. Therefore, TDG usually performs better than the existing approaches. By also introducing 1-D grids to reduce the error due to the uniformity assumption, HDG can give a significant improvement over existing approaches.}

\mypara{Contributions}
To summarize, this paper makes the following contributions:
\begin{itemize}
	\item We propose TDG and HDG for answering multi-dimensional range queries under LDP, which include a guideline for choosing the grid granularities based on analysis of errors from different sources.
	\item We conduct extensive experiments to evaluate the performance of different approaches using both real and synthetic datasets. The results show that HDG outperforms existing approaches by one order of magnitude.
\end{itemize}

\mypara{Roadmap} \rv{Section~\ref{sec:back} provides the preliminaries. Section~\ref{sec:baseline_approaches} describes the problem statement and four baseline approaches.}
Section~\ref{sec:grid_approaches} gives the details of our grid approaches. Section~\ref{sec:experiments} shows our experimental results. Section~\ref{sec:related}
reviews related work. Finally, Section~\ref{sec:conc} concludes this paper.

\section{Preliminaries}
\label{sec:back}

\subsection{Local Differential Privacy}
\label{subsec:ldp}

Local differential privacy (LDP)~\cite{DBLP:conf/focs/KasiviswanathanLNRS08} offers a high level of privacy protection, since each user only reports the sanitized data. Each user's privacy is still protected even if the aggregator is malicious.  In particular, each user perturbs the value $v$ using a randomized algorithm $\AA$ and reports $\AA(v)$ to the aggregator. Formally, LDP is defined in the following.
\begin{definition}[Local Differential Privacy] \label{def:ldp}
	An algorithm $\AA(\cdot)$ satisfies $\varepsilon $-local differential privacy ($\varepsilon$-LDP), where $\varepsilon \geq 0$,
	if and only if for any pair of inputs $(v, v')$, and any set \results of possible outputs of $\AA$, we have
	\begin{equation*}
		\Pr{\AA(v)\in \results} \leq e^{\varepsilon}\, \Pr{\AA(v')\in \results}.
	\end{equation*}
\end{definition}

\subsection{Categorical Frequency Oracles}
\label{subsec:fo}

In LDP, most problems can be reduced to frequency estimation. Below we present two state-of-the-art Categorical Frequency Oracle (CFO) protocols for these problems.

\mypara{Randomized Response}
\label{subsubsec:rr}
The basic protocol in LDP is random response~\cite{10.2307/2283137}.  It was introduced for the binary case, but can be easily generalized to the categorical setting.  Here we present the generalized version of random response (\grr), which enables the estimation of the frequency of any given value in a fixed domain.

Here each user with value $v \in [c]$ sends the true value $v$ with probability $p$, and with probability $1-p$ sends a randomly chosen $v'\in [c]$ s.t. $v'\ne v$.  More formally, the perturbation function is defined as
\begin{align}
	\forall_{y \in [c]}\;\Pr{\grr(v) = y}  = \left\{
	\begin{array}{lr}
		p=\frac{e^\varepsilon}{e^\varepsilon + c - 1}, & \mbox{if} \; y = v    \\
		p'= \frac{1}{e^\varepsilon + c - 1},           & \mbox{if} \; y \neq v \\
	\end{array}\label{eq:grr}
	\right.
\end{align}
This satisfies $\epsilon$-LDP since $\frac{p}{p'}=e^\varepsilon$.
To estimate the frequency of $\eef_v$ for $v\in [c]$, one counts how many times $v$ is reported, denoted by $\sum_{i\in[n]}\ind{y_i=v}$, and then computes
\begin{align*}
	\rv{\eef_v} =  \frac{1}{n}\sum_{i\in[n]}\frac{\ind{y_i=v}-p'}{p-p'},
\end{align*}
where $\ind{y_i=v}$ is the indicator function that the report $y_i$ of the $i$-th user  equals $v$, and $n$ is the total number of users.

Because each report $y_i$ is an independent random variable, by the linearity of variance, we can show that the variance for this estimation is
\begin{align}
	\Var{\rv{\eef_v}} = \frac{c-2+e^\varepsilon}{(e^\varepsilon-1)^2 \cdot n}.\label{eq:var_grr}
\end{align}

\mypara{Optimized Local Hash}
The optimized local hash (OLH) protocol deals with a large domain by first using a hash function to compress the input domain $[c]$ into a smaller domain $[c']$, and then applying randomized response to the hashed value.  In this protocol, both the hashing step and the randomization step result in information loss. The choice of the parameter $c'$ is a trade-off between loss of information during the hashing step and loss of information during the randomization step.  It is shown in~\cite{DBLP:conf/uss/WangBLJ17} that the estimation variance as a function of $c'$ is minimized when $c'=e^\varepsilon+1$.

In \olh, one reports $\tuple{H,\grr(H(v))}$ where $H$ is randomly chosen from a family of hash functions that hash each value in $[c]$ to a new one in $[c']$, and $\grr(\cdot)$ is the perturbation function for random response, while operating on the domain $[c']$ (thus $p = \frac{e^\varepsilon}{e^\varepsilon+c'-1}$ in Equation~\eqref{eq:grr}).
Let $\tuple{H_i,y_i}$ be the report from the $i$-th user.
For each value $v\in [c]$, to compute its frequency, one first computes $|\{i\mid H_i(v) = y_i\}| = \sum_{i\in[n]}\ind{ H_i(v) = y_i}$,
and then transforms it to its unbiased estimation
\begin{align*}
	\rv{\eef_v} =\frac{1}{n}\sum\limits_{i \in [n]} \frac{{ {\ind{{H_i}(v) = {y_i}} - 1/c'} }}{{p - 1/c'}}.
\end{align*}

In~\cite{DBLP:conf/uss/WangBLJ17}, it is shown that the estimation variance of \olh is
\begin{align}
	\Var{\rv{\eef_v}}=\frac{4e^\varepsilon}{(e^\varepsilon-1)^2\cdot n}. \label{eq:var_olh}
\end{align}
Compared with \grr, \olh has a variance that does not depend on $c$.
As a result, for a small $c$ (such that $c-2<3e^\varepsilon$), \grr is better; but for a large $c$, \olh is preferable.

\subsection{Principle of Dividing Users}
\label{subsec:sampling_technology}
Dividing users is one common feature among the existing LDP works~\cite{DBLP:journals/pvldb/CormodeKS19,DBLP:conf/sigmod/WangDZHHLJ19,DBLP:conf/ccs/ZhangWLHC18}. That is, when multiple pieces of information are needed, the best results are obtained by dividing users into groups, and then gathering information from each group. This is different from the traditional DP setting~\cite{DBLP:conf/tcc/DworkMNS06}, where there is a trusted aggregator having access to raw data records. In DP setting, the privacy budget is split to measure them all.  This is because the estimation variance in LDP setting is linear in the number of users, while in DP setting, it is a constant.  As a result, dividing users into $m$ groups incurs a $m^2$ multiplicative factor in DP setting (because the result is multiplied by $m$), while in LDP setting, this factor is only $m$ (because the number of users is divided by $m$).  As splitting privacy budget by $m$ increases variances for both cases by $m^2$, one prefers dividing users in LDP setting while splitting privacy budget in DP setting (as there is no sampling error). We will also apply this principle of dividing users in our proposed approaches.

\section{Problem Statement and Baseline Approaches}
\label{sec:baseline_approaches}

\subsection{Problem Statement}
\label{subsec:problem}

Consider there are $d$ ordinal attributes $\{ {a_1},{a_{2,}} \cdots ,{a_d}\} $.
Without loss of generality, we assume that all attributes have the same domain $[c]=\{1, 2,  \ldots, c\}$, where $c$ is a power of two (if not in real setting, we can simply add some dummy values to achieve it).
Let $n$ be the total number of users. The $i$-th user's record is a $d$-dimensional vector, denoted by $\mathbf{v}_i=\langle v_i^1, v_i^2,\ldots, v_i^d\rangle$ where $\rv{v_i^t}$ means the value of attribute $\rv{a_t}$ in record $\mathbf{v}_i$.

We focus on the problem of answering multi-dimensional range queries under LDP. In particular, a multi-dimensional range query is a conjunction of multiple predicates for the attributes in its interest.
\rv{Formally, a $\lambda$-dimensional ($\lambda$-D) range query $q$ is defined as
	\begin{equation*}
		q = ({a_{{t_1}}},[{l_{{t_1}}},{r_{{t_1}}}]) \wedge ({a_{{t_2}}},[{l_{{t_2}}},{r_{{t_2}}}]) \wedge  \cdots  \wedge ({a_{{t_\lambda }}},[{l_{{t_\lambda }}},{r_{{t_\lambda }}}]),
	\end{equation*}
	where $1\le t_\phi \le d$, and $t_\phi \ne t_\psi$ when $ \phi \ne \psi$.  We define $A_q$ to be $\{ a_{t_\phi} | 1\le \phi \le \lambda\}$ representing the set of attributes in $q$'s interest. Intuitively, such a query $q$ selects all records whose value of attribute $a_{t_\phi}$ is in the interval $[{l_{t_\phi}},{r_{t_\phi}}]$ for all $a_{t_\phi} \in A_q$.
	The answer of the query $q$ equals the fraction of these selected records. In particular, the real answer of $q$ can be represented as}
\begin{align*}
	\rv{\rf_q} = & \frac{|\{ \mathbf{v}_i \mid v_i^t \in [l_t, r_t], \forall {a_t} \in {A_q}\}|}{n}.\label{eq:query}
\end{align*}

In our problem setting, we assume that there is an aggregator that does not have access to the users' raw records. Our goal is to design an approach to enable the aggregator to get the answers of all possible range queries from the $n$ users while satisfying LDP.
Please see Table~\ref{tbl:notation} for the list of notations.

\mypara{\rv{Key Technical Challenges}}
\rv{To address this problem, we identify three key technical challenges:
	1) capturing the correlations among attributes,
	2) avoiding the curse of dimensionality, and
	3) coping with the large domains of attributes.
	Failure to solve any of these three challenges will lead to poor utility of the results.}

\vspace{3pt}
\rv{In the following, we will describe four baseline approaches that may handle the problem of answering multi-dimensional range queries under LDP and analyze how they deal with these challenges. In particular, the first two approaches CALM and HIO are existing approaches that can be directly applied to this problem. The third approach Low-dimensional HIO (LHIO) is an improvement of HIO.
	The last approach Multiplied Square Wave (MSW) is an extension of the existing approach that may answer 1-D range queries.}

\begin{table}
	\centering
	\begin{tabular}{@{}c|c@{}}
		\toprule
		\textbf{Notation} & \textbf{Meaning}                             \\ \midrule
		$n$               & The total number of users                    \\
		$d$               & The number of attributes                     \\
		$c$               & The domain size of an attribute              \\
		$b$               & The branching factor of a hierarchy          \\
		$m$               & The number of user groups                    \\
		$g$               & The granularity for an ordinal domain        \\
		$q$               & The range query                              \\
		\rv{$A_q$}        & \rv{The set of attributes in $q$'s interest} \\
		$\lambda$         & The query dimension                          \\

		\bottomrule
	\end{tabular}
	\caption{Notations}
	\vspace{-9pt}
	\label{tbl:notation}
\end{table}

\subsection{CALM}
\label{subsec:calm}

\vspace{3pt}
CALM~\cite{DBLP:conf/ccs/ZhangWLHC18} is the state-of-the-art for marginal release under LDP. In particular, a $\lambda$-D marginal means the joint distribution of $\lambda$ attributes. Due to the curse of dimensionality, directly computing a high-dimensional marginal using a LDP frequency oracle will lead to too much added noise. To solve this problem, CALM proposes to collect low-dimensional marginals and reconstruct a high-dimensional marginal from them.
We notice that CALM can be used to answer range queries. In particular, for a $\lambda$-D range query, one can employ CALM to get its answer by directly summing up the noisy marginals included in the query.

\rv{CALM only captures necessary pair-wise attribute correlations, which effectively overcomes the first two challenges.
	However, it fails to solve the third challenge. To answer a range query, CALM needs to sum up all noisy marginals in the query, which may result in a large amount of noise in the answer when $c$ is relatively large.}

\subsection{HIO}
\label{subsec:hio}

HIO~\cite{DBLP:conf/sigmod/WangDZHHLJ19} is a hierarchy-based approach that can directly answer multi-dimensional range queries under LDP. In HIO, given $d$ attributes with the domain $[c]$, the aggregator first constructs a 1-D hierarchy for each attribute.
To be specific, a 1-D hierarchy is a hierarchical collection of intervals with a branching factor $b$. The root corresponds to the entire domain $[c]$ and is recursively partitioned into $b$ equally sized subintervals until the leaves whose corresponding subintervals only contain one value are reached. Thus there are $h = \log_b c+1$ levels, called one-dim levels, in a 1-D hierarchy.
By defining that the root has a level 0, there are $b^{\ell}$ subintervals in a level $\ell \in [0, h] $. It is found in~\cite{DBLP:conf/sigmod/WangDZHHLJ19} that the optimal $b$ is around $5$.
For illustration, we define a $d$-dim level as a group of $d$ one-dim levels $(\ell_1, \ell_2, \ldots, \ell_d)$, each of which comes from one of these $d$  1-D hierarchies.
Similarly, we define a $d$-dim interval as a group of $d$ intervals, each of which also comes from one of these $d$ 1-D hierarchies.

Then, the aggregator constructs a $d$-dimensional hierarchy with these $d$ 1-D hierarchies.
A level in the $d$-dimensional hierarchy is actually a $d$-dim level. Thus there are $(h+1)^d$ $d$-dim levels in the $d$-dimensional hierarchy.
Since there are $b^{\ell}$ subintervals in a one-dim level ${\ell}$ in a 1-D hierarchy, a $d$-dim level $(\ell_1, \ell_2, \ldots, \ell_d)$ includes $\prod\limits_{i = 1}^d {{b^{{\ell _i}}}}$ $d$-dim intervals.
Next, the aggregator randomly divides users into $(h+1)^d$ groups, where each group reports one $d$-dim level. After using OLH to get the noisy frequencies of all $d$-dim intervals in every $d$-dim level, the aggregator can answer a multi-dimensional range query in the following manner.

To answer a $\lambda$-D range query
\begin{equation*}
	\rv{q = ({a_{{t_1}}},[{l_{{t_1}}},{r_{{t_1}}}]) \wedge ({a_{{t_2}}},[{l_{{t_2}}},{r_{{t_2}}}]) \wedge  \cdots  \wedge ({a_{{t_\lambda }}},[{l_{{t_\lambda }}},{r_{{t_\lambda }}}]),}
\end{equation*}
the aggregator first expands $q$ to a new $d$-dimensional range query $q'$ that is interested in all $d$ attributes by assigning a specified interval $[1, c]$ for each attribute not in $A_q$.
Then, for each attribute in these $d$ attributes, the aggregator finds the least number of subintervals that can make up its specified interval in $q'$ from its corresponding 1-D hierarchy. Finally, the aggregator sums up the noisy frequencies of all the $d$-intervals consisting of them to get the answer of $q'$, which is equivalent to that of $q$.

\rv{HIO solves the first challenge by capturing the correlations among all attributes. However, HIO fails to handle the other two challenges. In HIO, users are divided into $(h+1)^d$ groups where $h = \log_b c$.  When $d$ or $c$ is relatively large, there are too few users in each group, which will incur a high magnitude of added noise in the frequencies of $d$-dim intervals and result in large errors.}

\vspace{3pt}
\subsection{LHIO: Low-dimensional HIO}
\label{subsec:lhio}

We observe that CALM achieves good utility by using 2-D marginals to reconstruct high-dimensional ones. Using this idea, we can modify HIO, resulting in a new approach called Low-dim HIO (LHIO). Its main idea is to compute the answers of 2-D range queries and then estimate the answer of a high-dimensional range query from them.
Specifically, the aggregator first generates all $\binom{d}{2}$ attribute pairs from the given $d$ attributes and then randomly divides users into $m =\binom{d}{2}$ groups, where each group works on one pair of attributes. Next, for each attribute pair, the aggregator invokes HIO to construct a 2-D hierarchy by interacting with its corresponding user group. The constructed 2-D hierarchies can be directly used to answer all possible 2-D range queries. To estimate the answer of a higher dimensional range query, the aggregator invokes the estimation method which will be presented in Section~\ref{subsec:estimation_for_lambda_query}.

However, directly using the obtained noisy frequencies will lead to two inconsistency problems in our setting. The first one is within a 2-D hierarchy. That is, different levels of the noisy hierarchy may give inconsistent estimations due to LDP noise. The second one is among different 2-D hierarchies. Since each attribute is related to $d-1$ pairs, the frequencies marginalized on it from these $d-1$ 2-D hierarchy are usually different. The accuracy of the answers of the 2-D range queries will increase if the problem can be solved.
We identify that the key to remove inconsistency is to solve the first inconsistency problem, since the second one can be easily solved by the overall consistency in CALM after the first one is handled.
Therefore, we focus on the first problem and develop a new method to enforce consistency within a 2-D hierarchy. Its main idea is to adapt the constrained inference in Hay et al.~\cite{DBLP:journals/pvldb/HayRMS10} to a 2-D hierarchy and perform the operation twice by starting with the first and second attribute of the attribute pair, respectively. Its details are omitted due to space limitation.

LHIO satisfies $\varepsilon$-LDP because the report from each user uses OLH and satisfies $\varepsilon$-LDP. We show that by avoiding directly handling high-dimensional queries and removing inconsistency, LHIO can perform much better than HIO.
\rv{Similar to CLAM, LHIO overcomes the first two challenges by capturing necessary pair-wise attribute correlations. However, LHIO fails to solve the third challenge. In LHIO, users are divided into $\binom{d}{2} \cdot (h+1)^2$ groups where $h = \log_b c$. For a relatively large $c$, it will also bring about excessive noises.}

\subsection{MSW: Multiplied Square Wave}
\label{subsec:multiplied_square_wave}

Recently, Li et al.~\cite{DBLP:conf/sigmod/Li0LLS20} proposed an approach called Square Wave (SW) for estimating the distribution of a single numerical attribute under LDP.
It takes advantage of the ordinal nature of the domain and reports values that are close to the true value with higher probabilities than values that are farther away from the true value.

For handling an attribute with the discrete domain $[c]$, we initially normalize it to the continuous domain $[0,1]$.
Given a value $v \in [0, 1]$, SW perturbs it as:
\begin{align}
	\forall y \in [-\delta,1+\delta],  \;\Pr{\mbox{SW}(v)=y}  \!=\! \left\{
	\begin{array}{lr}
		\!p,  & \mbox{if} \; |v-y|\le \delta  \ , \\
		\!p', & \mbox{ otherwise} \ ,             \\
	\end{array}\nonumber
	\right.
\end{align}
where $\delta  = \frac{\varepsilon e^\varepsilon - e^\varepsilon +1}{2e^\varepsilon(e^\varepsilon - 1 - \varepsilon)}$ is the ``closeness'' threshold.
By maximizing the difference between $p$ and $p'$ while satisfying that the total probability adds up to $1$, the values $p$ and $p'$ can be derived as $p =  \frac{e^\varepsilon}{2 \delta e^\varepsilon + 1}$ and $p' =  \frac{1}{2\delta e^\varepsilon + 1}$, respectively.
After receiving perturbed reports from all users, the aggregator runs the Expectation Maximization algorithm to find an estimated distribution that maximizes the expectation of the observed output. It is shown in~\cite{DBLP:conf/sigmod/Li0LLS20} that SW outperforms other approaches for answering 1-D range queries.

Here we introduce Multiplied Square Wave (MSW), which is extended from SW to handle multi-dimensional range queries under LDP.
In MSW, given $d$ attributes, the aggregator randomly divides users into $d$ groups, where each group reports one attribute.  After utilizing SW to obtain the distribution of each individual attribute, a multi-dimensional range query is answered by using the product of the answers of all associated 1-D range queries. Such approximation implicitly assumes that all attributes are independent.

\vspace{3pt}
\rv{In MSW, each user only reports one attribute via SW that satisfies $\varepsilon$-LDP. Therefore, this process can ensure $\varepsilon$-LDP for each user.
	In addition, the subsequent multiplication post-process steps take those outputs that are already differentially private and does not access any user's raw data. Thus, MSW satisfies $\varepsilon$-LDP.}
\rv{Since MSW only collects the information of individual attributes, it solves the last two challenge. However, it fails to handle the first challenge. MSW totally loses the correlations among attributes, which will produce high errors when handling correlated attributes.}

\section{Grid Approaches}
\label{sec:grid_approaches}

\rv{In this section, we first elaborate our grid approaches for answering multi-dimensional range queries under LDP in Section~\ref{subsec:overview}-\ref{subsec:estimation_for_lambda_query}.
	Then we give their privacy and utility analysis in Section~\ref{subsec:theoretical_analysis}.
	Finally, we describe how to choose the proper granularities in Section~\ref{subsec:choosing_the_granularity}.}

\subsection{Overview}
\label{subsec:overview}

As analysed in \rv{Section~\ref{sec:baseline_approaches}}, none of the baseline approaches can overcome all three key challenges.
To address this problem, we first propose an approach called Two-Dimensional Grid (TDG).
Its main idea is to carefully use binning to partition the 2-D domains of all attribute pairs into 2-D grids that can answer all 2-D range queries and then estimate the answer of a higher dimensional range query from the answers of the associated 2-D range queries.

However, since values within the same cell in a grid are reported together, the aggregator cannot tell the distribution within each cell and only assumes a uniform distribution. When computing the answer of a 2-D range query by the cells that are partially included in the query, this may lead to large error due to the uniformity assumption. To correct this deficiency, we further propose an upgraded approach called Hybrid-Dimensional Grid (HDG), which also introduces finer-grained 1-D grids and combines the information from 1-D and 2-D grids to answer range queries.

\rv{Note that the first two challenges pose a dilemma: capturing full correlations (as HIO) will lead to the curse of dimensionality; while only focusing on individual attributes (as MSW) will totally lose correlation information. In CALM~\cite{DBLP:conf/ccs/ZhangWLHC18}, the similar dilemma is solved by using 2-D marginals to reconstruct high-dimensional ones, which achieves a good trade-off when handling these two challenges. Inspired by this idea, both TDG and HDG choose to capture the necessary pair-wise attribute correlations via 2-D grids, which overcomes the first two challenges. The third challenge is also carefully solved in TDG and HDG by properly using binning with the guideline to reduce the error incurred by a large domain.}

Specifically, both TDG and HDG consist of three phases:

\vspace{3pt}
\textbf{Phase 1. Constructing Grids.} In TDG, from the given $d$ attributes, the aggregator first generates all $\binom{d}{2}$ attribute pairs. Then the aggregator randomly divides users into $m = \binom{d}{2}$ groups, each of which corresponds to one pair. Next, for each attribute pair $(a_j, a_k)$ where $1 \le j < k \le d$, the aggregator assigns the same granularity $g_2$ to construct a 2-D grid $G^{(j, k)}$ by partitioning the 2-D domain $\rv{[c] \times [c]}$ into $g_2 \times g_2$ \rv{2-D} cells of equal size. \rv{In particular, each 2-D cell specifies a 2-D subdomain consisting of $\frac{c}{g_2} \times \frac{c}{g_2}$ 2-D values.} Finally, to obtain noisy frequencies of cells in each grid, the aggregator instructs each user in the group corresponding to the grid to report which cell his/her private value is in using OLH.

In HDG, the aggregator also constructs $d$ 1-D grids for the $d$ attributes, respectively. Thus there will be  $d + \binom{d}{2}$ grids in HDG and users are divided into $m = d + \binom{d}{2}$ groups, each of which corresponds to one of these grids. In addition to constructing $\binom{d}{2}$ 2-D grids with granularity $g_2$ as TDG, \rv{in HDG}, the aggregator assigns the identical granularity $g_1$ to construct a 1-D grid $G^{(j)}$ containing $g_1$ \rv{1-D} cells of equal size for each single attribute $a_j (1 \le j \le d)$. \rv{In particular, each 1-D cell specifies a 1-D subdomain consisting of $\frac{c}{g_1}$ 1-D values.}  Finally, as in TDG, the aggregator uses OLH to obtain noisy frequencies of cells in each grid.

\vspace{3pt}
\textbf{Phase 2. Removing Negativity and Inconsistency.} Due to using OLH to ensure privacy, the noisy frequency of a cell may be negative, which violates the prior knowledge that the true one is non-negative. Moreover, since an attribute is related to multiple grids, the noisy frequencies integrated on the attribute in different grids may be different, leading to inconsistency among grids. In this phase, to improve the utility, the aggregator post-processes the constructed grids to remove the negativity and inconsistency. The difference between TDG and HDG is that TDG only requires the aggregator to handle 2-D grids while 1-D and 2-D grids needs to be handled together in HDG. We describe the detail for post-processing grids in Section~\ref{subsec:consistency_operation}.

\vspace{3pt}
\textbf{Phase 3. Answering Range Queries.}
In this phase, the aggregator can answer all multi-dimensional range queries. We first describe how to answer a 2-D range query. \rv{For ease of illustration, we take a 2-D range query $q_0$ interested in $A_{q_0} = \{a_1, a_2\}$ as an example. In TDG, to get the answer $\eef_{q_0}$ of $q_0$, the aggregator first finds the 2-D grid $G^{(1,2)}$ corresponding to $A_{q_0}$ and then checks all \rv{2-D} cells in $G^{(1,2)}$ in the following manner.} If a cell is completely included in \rv{$q_0$}, the aggregator includes its noisy frequency in \rv{$\eef_{q_0}$}; if a cell is partially included, the aggregator estimates the sum of frequencies of common \rv{2-D} values between the cell and $\rv{q_0}$ by uniform guess, i.e., assuming that the frequencies of \rv{2-D} values within the cell are uniformly distributed and then adds the sum to $\rv{\eef_{q_0}}$.

\rv{In HDG, the aggregator treats those cells partially included in $q_0$ using a response matrix rather than uniform guess, which can significantly improve the accuracy of results. To be specific, for each attribute pair $(a_j, a_k)$, the aggregator first employs the three grids $\{G^{(j)}, G^{(k)}, G^{(j, k)}\}$ to build a response matrix $M^{(j, k)}$ before answering 2-D range queries. In particular, the matrix $M^{(j, k)}$ consists of $c \times c$ elements that are in one-to-one correspondence with the estimated frequencies of 2-D values in the 2-D domain $\rv{[c] \times [c]}$ of $(a_j, a_k)$.
	The details of response matrix generation are given in Section~\ref{subsec:response_matrix_generation}.
	When calculating the answer $\eef_{q_0}$ of the 2-D query $q_0$ in HDG, the aggregator also checks all 2-D cells in the grid $G^{(1,2)}$ corresponding to $A_{q_0}$. For a cell completely included in $q_0$, the aggregator includes its noisy frequency in \rv{$\eef_{q_0}$} as in TDG; for a cell partially included in the query $q_0$, the aggregator identifies the common 2-D values between this cell and $q_0$, and then adds the sum of their corresponding elements in $M^{(1,2)}$ to $\eef_{q_0}$.}

For a $\lambda$-D range query where $\lambda > 2$, its answer cannot be directly obtained from the constructed \rv{2-D} grids or response matrices. \rv{To answer this $\lambda$-D query,  we propose to split it into $\binom{\lambda}{2}$ associated 2-D range queries and then estimate its answer from all answers of these $\binom{\lambda}{2}$ 2-D queries.} We discuss it in detail in Section~\ref{subsec:estimation_for_lambda_query}.

\subsection{Post-Process for Grids}
\label{subsec:consistency_operation}

The post-process for grids contains two basic steps including non-negativity step and consistency step, which are used to remove negativity and inconsistency, respectively.

\vspace{2pt}
\textbf{Non-Negativity Step.}
In this step, the aggregator handles the estimated frequencies of cells in each grid by Norm-Sub~\cite{DBLP:journals/corr/abs-1905-08320}, which can make all estimates non-negative and sum up to 1. In Norm-Sub, firstly, all negative estimates are converted to 0. Then the total difference between 1 and the sum of positive estimates is calculated. Next, the average difference is obtained through dividing the total difference by the number of positive estimates. Finally, every positive estimate is updated by subtracting the average difference. The process is repeated until all estimates become non-negative.

\vspace{2pt}
\textbf{Consistency Step.} We first describe how to achieve consistency on an attribute among grids. For an attribute $a$, it is related to $d$ grids in total, which includes one 1-D grid and $d-1$ 2-D grids. Assume these $d$ grids are  $\{{G_1},{G_2}, \cdots {G_d}\}$.
\rv{For an integer $j \in [1, {g_2}]$, we define ${{\rm P}_{{G_i}}}(a,j)$ to be the sum of  frequencies of $G_i$'s cells whose specified subdomain corresponds to $a$ is in $[(j - 1) \times \frac{c}{{{g_2}}} + 1,j \times \frac{c}{{{g_2}}}]$.}
To make all ${{\rm P}_{{G_i}}}(a,j)$ consistent,  we compute their weighted average as
${\rm P}(a,j) = \sum\limits_{i = 1}^d {{\theta _i} \cdot {{\rm P}_{{G_i}}}(a,j)}$,
where $\theta_i$ is the weight of ${{{\rm P}_{{G_i}}}(a,j)}$.

To get a better estimation, we need to carefully set the value of $\theta_i$. Our goal is to minimize the variance of ${\rm P}(a,j)$, i.e. $\Var{{\rm P}(a,j)} = \sum\limits_{i = 1}^d {{\theta _i}^2 \cdot \Var{{{\rm P}_{{G_i}}}{(a,j)}} } = \sum\limits_{i = 1}^d {{\theta _i}^2 \cdot |{S_i}| \cdot {\mathsf{Var}_0}}$, where $S_i$ is the set of cells whose frequencies contribute to ${{{\rm P}_{{G_i}}}(a,j)}$ and ${\mathsf{Var}_0}$ is the basic variance for estimating a single cell (we assume each user group has the same population).
Apparently, if $G_i$ is 1-D, ${S_i} = \frac{{{g_1}}}{{{g_2}}}$; if $G_i$ is 2-D, ${S_i} = {g_2}$.
Based on the analysis in~\cite{DBLP:conf/ccs/ZhangWLHC18}, we have ${\theta _i} = {\frac{1}{{|{S_i}|}}} / {{\sum\limits_{i = 1}^d {\frac{1}{{|{S_i}|}}} }}$ and
the optimal weighted average is ${\rm P}(a,j) = \left( {\sum\limits_{i = 1}^d {\frac{1}{{|{S_i}|}} \cdot {{\rm P}_{{G_i}}}(a,j)} } \right)  / {{\sum\limits_{i = 1}^d {\frac{1}{{|{S_i}|}}} }}$.
Once the ${\rm P}(a,j)$ is obtained, we need to make each ${{{\rm P}_{{G_i}}}(a,j)}$ equal it, which can be achieved in the following manner. For each cell in $S_i$, we update its frequency by adding the amount of change $\left ({{\rm P}(a,j) - {{\rm P}_{{G_i}}}(a,j)} \right) / {|{S_i}|}$.

To achieve consistency among all attributes, we can use the above method one by one for each single attribute. It is shown in~\cite{DBLP:conf/sigmod/QardajiYL14} that following any order of these attributes, a later consistency step will not invalidate consistency established in previous steps.

\vspace{3pt}
Note that applying the consistency step may incur negativity, and vise versa. Thus in the post-process, we interchangeably invoke these two steps multiple times. Since we need to ensure non-negativity for the response matrix generation in Phase 3, we end the post-process with the non-negativity step. While the last step may again introduce inconsistency, it tends to be very small.

\subsection{Response Matrix Generation}
\label{subsec:response_matrix_generation}

\rv{For an attribute pair $(a_j, a_k)$, it corresponds to the response matrix $M^{(j, k)}$ of size $c \times c$, where the element $M^{(j, k)}[\beta_j, \beta_k]$ represents the estimated frequency of 2-D value $(\beta_j, \beta_k)$ in the $[c] \times [c]$ 2-D domain of $(a_j, a_k)$. To build $M^{(j, k)}$, we propose to invoke the efficient estimation method Weighted Update~\cite{DBLP:journals/toc/AroraHK12,DBLP:conf/nips/HardtLM12} with the three grids $\{G^{(j)}, G^{(k)}, G^{(j, k)}\}$ corresponding to $\{a_j, a_k, (a_j, a_k)\}$, respectively. Its main idea is to keep using the information on each cell in these three grids to update the matrix until each cell's frequency equals the sum of its corresponding elements in the matrix.}

\rv{Algorithm~\ref{alg:get_estimation_matrix} provides the details of building response matrix $M^{(j, k)}$ for attribute pair $(a_j, a_k)$. It takes grids $\{G^{(j)}, G^{(k)}, G^{(j, k)}\}$ and domain size $c$ as inputs and outputs the response matrix $M^{(j, k)}$.}
\rv{In Algorithm~\ref{alg:get_estimation_matrix}, for each grid $G$ in $\{G^{(j)}, G^{(k)}, G^{(j, k)}\}$, the aggregator performs the following update process on $M^{(j,k)}$.
	For each cell $s$ in $G$, the aggregator first finds the set of 2-D values $\crf(s)$ corresponding to $s$, which means that $\crf(s)$ consists of all those 2-D values whose frequency can contribute to the frequency $\eef_s$ of cell $s$.}
\rv{To illustrate the definition of $\crf(s)$, we take a 2-D cell $s$ in $G^{(j, k)}$ as an example. Assume the 2-D cell $s$ specifies a 2-D subdomain $[l_s^j, r_s^j] \times [l_s^k, r_s^k]$, where $[l_s^j, r_s^j]$ and $[l_s^k, r_s^k]$ correspond to $a_j$ and $a_k$, respectively. Then, $\crf(s)$ can be represented as
	\begin{equation*}
		\crf(s) = \{(\beta_j, \beta_k)| \beta_j \in [l_s^j, r_s^j], \beta_k \in [l_s^k, r_s^k]\}.
	\end{equation*}
	Note that this representation is also applicable to a 1-D cell $s$ in $G^{(j)}$ (or $G^{(k)}$), since we can equivalently transform its specified 1-D subdomain $[l_s^j, r_s^j]$ (or $[l_s^k, r_s^k]$) into 2-D subdomain $[l_s^j, r_s^j] \times [1, c]$ (or $[1, c] \times [l_s^k, r_s^k]$.).
	With $\crf(s)$, the aggregator updates the elements in $M^{(j,k)}$ as Lines 6-9 in Algorithm~\ref{alg:get_estimation_matrix}.
	This update process is repeated until convergence.}

\rv{In Algorithm~\ref{alg:get_estimation_matrix}, the convergence criteria is that the sum of the changes of all elements in the response matrix after each update process is lower than a given threshold. By comparing the results of setting different thresholds, we found that the results are almost the same so long as threshold is smaller than $\frac{1}{n}$.}

\subsection{Estimation for $\lambda$-D Range Query}
\label{subsec:estimation_for_lambda_query}

\begin{algorithm}[t]
	\caption{Building Response Matrix}
	\label{alg:get_estimation_matrix}
	\begin{algorithmic}[1]
		\Require{\rv{Grids $\{G^{(j)}, G^{(k)}, G^{(j, k)}\}$}, domain size $c$}
		\Ensure{\rv{Response matrix $M^{(j,k)}$}}
		\State \rv{initialize all $c \times c$ elements in the matrix $M^{(j, k)}$ as $\frac{1}{{{c^2}}}$;}
		\Repeat
		\rv{\For{each grid $G$ in $\{G^{(j)}, G^{(k)}, G^{(j, k)}\}$}
			\For{each cell $s$ in $G$}
			\State Find the set of 2-D values $\crf(s)$ corresponding to $s$;
			\State Calculate $\sums = \sum\limits_{(\beta_j, \beta_k) \in \crf(s)} M^{(j,k)}[\beta_j, \beta_k]$;
			\If{$\sums \ne 0$}
			\For{each 2-D value $(\beta_j, \beta_k)$ in $\crf(s)$}
			\State  $M^{(j,k)}[\beta_j, \beta_k] \gets \frac{M^{(j,k)}[\beta_j, \beta_k]}{\sums} \cdot \eef_s$;
			\EndFor
			\EndIf
			\EndFor
			\EndFor}
		\Until {convergence}
		\State \Return $\rv{M^{(j,k)}}$
	\end{algorithmic}
\end{algorithm}

\rv{To estimate the answer $\eef_q$ of a $\lambda$-D range query
	\begin{equation*}
		q = ({a_{{t_1}}},[{l_{{t_1}}},{r_{{t_1}}}]) \wedge ({a_{{t_2}}},[{l_{{t_2}}},{r_{{t_2}}}]) \wedge  \cdots  \wedge ({a_{{t_\lambda }}},[{l_{{t_\lambda }}},{r_{{t_\lambda }}}])
	\end{equation*}
	where $A_q=\{ a_{t_\phi} | 1\le \phi \le \lambda\}$,
	the aggregator first splits $q$ into $\binom{\lambda}{2}$ associated 2-D range queries
	\begin{equation*}
		\left\{q^{({j},{k})}=(a_j, [{l_j},{r_j}]) \wedge (a_k, [{l_k},{r_k}]) | a_j, a_k \in A_q \right\},
	\end{equation*}
	and then gets their answers $\left\{ {\eef_{q^{({j},{k})}}}\mid {{a_j},{a_k} \in {A_q}} \right\}$
	as described in Section~\ref{subsec:overview}.
	Finally, the aggregator uses these $\binom{\lambda}{2}$ 2-D queries' answers  to estimate ${\eef_q}$.}

\rv{In general, such an estimation problem can be solved by Maximum Entropy Optimization~\cite{DBLP:conf/sigmod/QardajiYL14,DBLP:conf/ccs/ZhangWLHC18}. (For self-containment, we include its description in Appendix~\ref{appendix:maximum_entropy}.)
	However, in experiments, we observe that Maximum Entropy Optimization cannot converge quickly in some cases. Therefore, we propose to use Weighted Update~\cite{DBLP:journals/toc/AroraHK12,DBLP:conf/nips/HardtLM12} to solve this estimation problem, which can achieve almost the same accuracy while with higher efficiency.}

\rv{Algorithm~\ref{alg:weighted_update} gives the procedure of estimating the answer of a $\lambda$-D range query $q$.
	It takes the answers of $\binom{\lambda}{2}$ associated 2-D queries as inputs and outputs a estimated answer vector $\mathbf{z}$. In particular, the vector $\mathbf{z}$
	consists of $2^\lambda$ elements that are in one-to-one correspondence with the answers of $2^\lambda$ $\lambda$-D queries in
	\begin{equation*}
		Q(q) = \{ { \wedge _t}({a_t},[{l_t},{r_t}] \text{ or } {[{l_t},{r_t}]' }) \mid {a_t} \in {A_q}\},
	\end{equation*}
	where the interval $[{l_t},{r_t}]'$ is the complement of $[l_t, r_t]$ on the domain of $a_t$.
	In Algorithm~\ref{alg:weighted_update}, for each ${\eef_{q^{({j},{k})}}}$ in $\left\{ {f_{q^{({j},{k})}}}\mid {{a_j},{a_k} \in {A_q}} \right\}$, the aggregator performs the following update process on $\mathbf{z}$.
	The aggregator first finds the set of $\lambda$-D queries ${Q(q)^{(j,k)}}$ corresponding to the 2-D query $q^{({j},{k})}$, which means that ${Q(q)^{(j,k)}}$ consists of all those $\lambda$-D queries whose answers can contribute to ${\eef_{q^{({j},{k})}}}$. In particular, ${Q(q)^{(j,k)}}$ contains $2^{\lambda-2}$ $\lambda$-D queries from ${Q(q)^{(j,k)}}$ and is defined as
	$\left \{ { \wedge _t}({a_t},[{l_t},{r_t}] \text{ or } {[{l_t},{r_t}]' })\wedge q^{({j},{k})} \mid {a_t} \in {A_q}/\{ {a_j},{a_k}\} \right\}$.
	Then, the aggregator calculates the sum $\sums$ of $\mathbf{z}[q']$ for all $q' \in {Q(q)^{(j,k)}}$, where $\mathbf{z}[q']$ is the element corresponding to the answer of $q'$.
	Next, the aggregator uses ${\eef_{q^{({j},{k})}}}$ to update the elements in $\mathbf{z}$ as Lines 6-8.
	This process is repeated until convergence. The estimated answer $\eef_q$ of the $\lambda$-D query $q$ equals its corresponding element in $\mathbf{z}$, i.e., $\mathbf{z}[q]$.}

\rv{In Algorithm~\ref{alg:weighted_update}, the convergence criteria is that the sum of the changes of all elements in the estimated vector after each update process is lower than a given threshold. We also found that the results are almost the same so long as threshold is smaller than $\frac{1}{n}$.}

\subsection{Privacy and Utility Analysis}
\label{subsec:theoretical_analysis}

\mypara{Privacy Guarantee}
We claim that both TDG and HDG satisfy $\varepsilon$-LDP because all the information from each user to the aggregator goes through OLH with $\varepsilon$ as privacy budget, and no other information is leaked.

\mypara{Error Analysis}
Below we analyze the expected squared error between the true query answer and the estimated answer. There are four kinds of errors: noise error, sampling error, non-uniformity error, and estimation error.

\begin{algorithm}[t]
	\caption{\rv{Estimating Answer of $\lambda$-D Range Query}}
	\label{alg:weighted_update}
	\begin{algorithmic}[1]
		\Require{\rv{Associated 2-D queries' answers $\left\{ {\eef_{q^{({j},{k})}}}\mid {{a_j},{a_k} \in {A_q}} \right\}$}}
		\Ensure{Estimated answer vector $\mathbf{z}$}
		\State initialize all $2^{\lambda}$ elements in the vector $\mathbf{z}$ as $\frac{1}{{{2^{\lambda}}}}$;
		\Repeat
		\rv{\For{each $\eef_{q^{({j},{k})}}$ in $\left\{ {\eef_{q^{({j},{k})}}}\mid {{a_j},{a_k} \in {A_q}} \right\}$}
			\State Find the set of queries ${Q(q)^{(j,k)}}$ corresponding to $q^{({j},{k})}$;
			\State Calculate ${\sums} = \sum\limits_{q' \in {Q(q)^{(j,k)}}} \mathbf{z}[q']$;
			\If{$\sums \ne 0$}
			\For{each query $q'$ in ${Q(q)^{(j,k)}}$}
			\State $\mathbf{z}[q'] \gets \frac{\mathbf{z}[q']}{\sums} \cdot \eef_{q^{({j},{k})}}$;
			\EndFor
			\EndIf
			\EndFor}
		\Until {convergence}
		\State \Return $\mathbf{z}$

	\end{algorithmic}
\end{algorithm}

\textit{Noise and Sampling Error.}
The noise error is due to the use of LDP frequency oracles. To satisfy LDP, one adds, to each cell, an independently generated noise, and these noises have the same standard deviation. When summing up the noisy frequencies of cells to answer a query, the noise error is the sum of the corresponding noises. As these noises are independently generated zero-mean random variables, they cancel each other out to a certain degree. In fact, because these noises are independently generated, the variance of their sum equals the sum of their variances. Therefore, the finer granularity one partitions the domain into, the more cells are included in a query, and the larger the noise error is. The sampling error is incurred by using cells' frequencies obtained from a user group to represent those obtained from the entire population, since the user group may have different distribution from the global one.

The noise and sampling errors can be quantified together.
\rv{Suppose the estimation is run on a sample $\sds$ of the dataset $D$.  We use $\eef_v(X)$ and $\rf_v(X)$ to denote the estimated and true frequencies of $v$ in $X$, respectively.  For simplicity, the frequency on the original dataset $\rf_v(D)$ is written as $\rf_v$.} The expected squared error for estimating one value is
\rv{\begin{small}
		\begin{align}
			\ep{\left(\eef_v(\sds) - \rf_v\right)^2}\nonumber = & \ep{\left(\eef_v(\sds) - \rf_v(\sds)\right)^2} + \ep{\left(\rf_v(\sds) - \rf_v\right)^2}+\nonumber \\
			                                                    & 2\ep{(\eef_v(\sds) - \rf_v(\sds))\cdot(\rf_v(\sds) - \rf_v)}. \label{equ:wfo:sample:varproof}
		\end{align}
	\end{small}}
Specifically, Equation~\eqref{equ:wfo:sample:varproof} consists of three parts. The first part is the variance of frequency oracle, i.e.,
\rv{\begin{small}
		\begin{align}
			\ep{\left(\eef_v(\sds) - \rf_v(\sds)\right)^2} \nonumber
			= & \, m\cdot \frac{p'(1-p')}{n(p-p')^2} + m\cdot \frac{{\rf}_v(p-p')(1-p-p')}{n(p-p')^2}\nonumber.
		\end{align}
	\end{small}}
In the case of \olh, we have $p=1/2$, $p'=1/(e^\varepsilon+1)$, and the quantity equals $\frac{{4m{e^\varepsilon }}}{{n{{({e^\varepsilon } - 1)}^2}}}+\frac{m}{n} \cdot \rf_v$.

\rv{The second part is $\ep{\left(\rf_v(\sds) - \rf_v\right)^2} = \, \frac{m-1}{n-1} \rf_v (1 - \rf_v)$. And the third part is
	$2\ep{(\eef_v(\sds) - \rf_v(\sds))\cdot(\rf_v(\sds) - \rf_v)} = 0$. We observe that the second part is a constant which is much smaller than the first part. Ignoring the small factor $\frac{m}{n} \cdot \rf_v$ in the first part, the expected squared noise and sampling error can be dominated by $\frac{{4m{e^\varepsilon }}}{{n{{({e^\varepsilon -1)^2}}}}}$.
	Due to space limitation, we present the detailed derivation of the above equations in Appendix~\ref{appendix:detailed_derivation_of_equations}.}

\textit{Non-Uniformity Error.}
Non-uniformity error is caused by cells that intersect with the query rectangle, but are not contained in it. For these cells, we need to estimate how many data points are in the intersected cells assuming that the data points are uniformly distributed, which will lead to non-uniformity error when the data points are not uniformly distributed. The magnitude of this error in any intersected cell, in general, depends on the number of
data points in that cell, and is bounded by it. Therefore, the finer the partition granularity, the lower the non-uniformity error. Calculating precise non-uniformity error requires the availability of the true data distribution, which is not the case in our setting.  Thus we opt to compute the approximate non-uniformity error.

\textit{Estimation Error.} When estimating the answer of a $\lambda$-D range query where $\lambda > 2$ from the associated answers of 2-D range queries, estimation error will occur.
Since the estimation error is dataset dependent, there is no formula for estimating it.
In general, more accurate answers of 2-D range queries can result in a smaller estimation error. However, its feature that the magnitude is dependent on the dataset will introduce uncertainty, which means that an opposite result may appear in a few cases.

\subsection{Choosing Granularities}
\label{subsec:choosing_the_granularity}

Since the granularities $g_1, g_2$ can directly affect the utility of our gird approaches, we propose the following guideline for properly choosing them.

\textit{Guideline:} \rv{To minimize the sum of squared noise and sampling error and squared non-uniformity error,} the granularity $g_1$ for 1-D grids should be
${g_1} = \sqrt[3]{{\frac{{{n_1} \cdot {{({e^\varepsilon } - 1)}^2} \cdot {\alpha _1}^2}}{{2{m_1}{e^\varepsilon }}}}}$;
the granularity $g_2$ for 2-D grids should be computed as
${g_2} = \sqrt {2{\alpha _2} \cdot ({e^\varepsilon } - 1) \cdot \sqrt {\frac{{{n_2}}}{{{m_2}{e^\varepsilon }}}} }$,
where $\varepsilon$ is the total privacy budget, ${n_i} (i = 1,2)$ is the number of users used for $i$-D grids,  ${m_i} (i = 1,2)$ is the number of user groups for $i$-D grids, and $\{\alpha_1, \alpha_2\}$ are some small constants depending on the dataset.
\rv{For simplicity, we make each user group have the same population, i.e. $\frac{{{n_2}}}{{{m_2}}} = \frac{n}{{\binom{d}{2}}}$ for TDG  and $\frac{{{n_1}}}{{{m_1}}} = \frac{{{n_2}}}{{{m_2}}} = \frac{n}{{d + \binom{d}{2}}}$ for HDG.}
To ensure that $g_1$ and $g_2$ are divisible by domain size $c$ at the same time, for each of them, we take the power of two closest to its derived value as the final value. If the obtained granularity is larger than $c$, we set it to $c$ by default. Our experimental results suggest that setting $\alpha_1=0.7$ and $\alpha_2 =0.03$ can typically achieve good performance across different datasets.

\mypara{Analysis on $g_1$}
A range query on a 1-D grid specifies a query interval on the attribute corresponding to the grid. For an average case, we consider that the ratio of this interval to the attribute's domain size is $\frac{1}{2}$. When answering the query from a 1-D grid with granularity $g_1$, there are roughly $\frac{g_1}{2}$ cells included in this query.
\rv{The squared noise and sampling error is $\frac{{{g_1}}}{2} \cdot \frac{{4{m_1}{e^\varepsilon }}}{{{n_1}{{({e^\varepsilon } - 1)}^2}}} = \frac{{2{g_1}{m_1}{e^\varepsilon }}}{{{n_1}{{({e^\varepsilon } - 1)}^2}}}$.}

The non-uniformity error is proportional to the sum of frequencies of values in the cells that intersect with the two sides of the query interval.  Assuming that the non-uniformity error is $\frac{{{\alpha_1}}}{{{g_1}}}$ for some constant $\alpha_1$, then it has a squared error of ${\left( {\frac{{{\alpha _1}}}{{{g_1}}}} \right)^2}$.

\rv{The minimize the sum of the two squared errors}
$\frac{{2{g_1}{m_1}{e^\varepsilon }}}{{{n_1}{{({e^\varepsilon } - 1)}^2}}} + {\left( {\frac{{{\alpha _1}}}{{{g_1}}}} \right)^2}$,
we should set $g_1$ to $\sqrt[3]{{\frac{{{n_1} \cdot {{({e^\varepsilon } - 1)}^2} \cdot {\alpha _1}^2}}{{2{m_1}{e^\varepsilon }}}}}$.

\mypara{Analysis on $g_2$}
Here we extend the above analysis to the 2-D grid setting.
For a 2-D query, we assume that the ratio of each query interval to its corresponding attribute's domain size is $\frac{1}{2}$. \rv{Then the squared noise and sampling error is} ${(\frac{{{g_2}}}{2})^2} \cdot \frac{{4{m_2}{e^\varepsilon }}}{{{n_2}{{({e^\varepsilon } - 1)}^2}}} = \frac{{{\left( {{g_2}} \right)^2} \cdot {m_2}{e^\varepsilon }}}{{{n_2}{{({e^\varepsilon } - 1)}^2}}}$.

\def\metric{MNAE}
\def\setwidth{0.47}
\def\setheight{0.36}
\def\mpagewidth{0.48}

\def\setwidth{\fwidth}
\def\setheight{\fheight}

\def\atumlist{6}
\def\dslist{64}
\def\qdimlist{2, 4}
\def\dqv{0.5}
\def\dqvfile{5}
\def\qn{200}

\def\unstr{$n$}
\def\atnumstr{$d$}
\def\dqvstr{$\omega$}
\def\qdimstr{$\lambda$}
\def\qnstr{$|Q|$}
\def\epstr{$\varepsilon$}
\def\dsstr{$c$}

\foreach \folder in {all_algorithm_vary_epsilon}
	{
		\begin{figure*}[t]
			\centering
			\foreach \datasetname/\capdatasetname in {ipums2018/Ipums, bfive/Bfive, normal/Normal, laplace/Laplace}
				{
					\foreach \atnum in {6}
						{
							\foreach \ds in {64}
								{
									\begin{minipage}[t]{\mpagewidth\linewidth}
										\centering
										\foreach \qdim in \qdimlist
										{
											\subfigure[\textbf{\capdatasetname}, \qdimstr= \qdim]
											{
												\includegraphics[width=\setwidth\columnwidth, height=\setheight\columnwidth]{\figurepath{\folder}/dn-\datasetname-un-1000000-an-\atnum-ds-\ds-fot-2-pf-2-qn-\qn-qd-\qdim-dqv-\dqvfile-\metric.pdf}
											}
											\hspace{-0.1in}
										}
									\end{minipage}
								}
						}
				}
			\vspace{-0.1in}
			{\includegraphics[scale = 0.4]{\figurepath{\folder}/legend.pdf}}
			\caption{\rv{\textbf{Varying $\varepsilon$ on all datasets under setting of \unstr = $10^6$,  \atnumstr = $6$, \dsstr = $64$, \dqvstr = $0.5$, \qdimstr = $2, 4$. MAEs are shown in log scale.}}}
			\label{fig:comparison_different_approaches_varying_epsilon}
			\vspace{-4pt}
		\end{figure*}
	}

The non-uniformity error is proportional to the sum of the frequencies of values in the cells that fall on the four edges of the query rectangle.
The query rectangle's edges contain $4 \cdot \frac{{{g_2}}}{2} = 2{g_2}$ cells; and the expected sum of frequencies of values included in these cells is $2{g_2} \cdot \frac{1}{{{g_2} \times {g_2}}} = \frac{2}{{{g_2}}}$. Similar to the 1-D grid setting, we assume that the non-uniformity error on average is some portion of it. Then the
squared error from non-uniformity is ${\left( {\frac{{2{\alpha _2}}}{{{g_2}}}} \right)^2}$ for some constant $\alpha_2$.
\rv{Our goal is to select $g_2$ to minimize the sum of the two squared errors}
$\frac{{2{g_1}{m_1}{e^\varepsilon }}}{{{n_1}{{({e^\varepsilon } - 1)}^2}}} + {\left( {\frac{{{\alpha _1}}}{{{g_1}}}} \right)^2}$.
\rv{To achieve this goal, ${g_2}$ should be} $\sqrt {2{\alpha _2} \cdot ({e^\varepsilon } - 1) \cdot \sqrt {\frac{{{n_2}}}{{{m_2}{e^\varepsilon }}}} }$.

\mypara{\rv{Discussion}}
\rv{In the analysis of non-uniformity error, for a cell that contributes to this error, we calculate the expected sum of frequencies of values in this cell based on the uniformity assumption. Although this assumption may lead to the deviation between the calculated error and the true one, it helps the analysis become more general for diverse datasets. Moreover, since 1-D grids are  finer-grained, this deviation's influence on the performance of HDG tends to be negligible. Thus, such an assumption still makes our guideline consistently effective for HDG when handling diverse datasets.}
\rv{Note that the recommended values of $\{\alpha_1, \alpha_2\}$ are obtained by tuning them on synthetic datasets under different setting of $n, c, d$, which does not leak any real users' private information. Besides, all other needed parameters for choosing granularities are derived from public background knowledge and do not require the aggregator to access raw data. Therefore, configuring TDG and HDG with our guideline will not lead to any privacy leakage.}

\section{Experimental Evaluation}
\label{sec:experiments}

In this section, we aim to answer the following questions: (1) how does our proposed HDG perform, (2) how can different parameters affect the results, and 3) how effective is the guidance for choosing granularities given by our guideline.

\subsection{Setup}\label{ssec:experimental_setup}

\mypara{Datasets}
We make use of two real datasets and two synthetic datasets in our experiments.
\begin{itemize}
	\item Ipums~\cite{data:ipums}: It is from the Integrated Public Use Microdata Series and has around 1 million records of the United States census in 2018.
	\item Bfive~\cite{data:Bfive}: It is collected through an interactive on-line personality test and contains around 1 million records. Each record describes the time spent on each question in milliseconds.
	\item Normal: This dataset is synthesized from multivariate normal distribution with mean 0, standard deviation 1. The covariance between every two attributes is 0.8.
	\item Laplace: This dataset is synthesized from multivariate laplace distribution with mean 0, standard deviation 1. The covariance between every two attributes is 0.8.
\end{itemize}

For the first two real datasets, we sample 1 million user records. To experiment with different numbers of users, we generate multiple test datasets from the two synthetic datasets with the number of users ranging from 100k to 10M. For evaluation varying different numbers of attributes and domain sizes, we generate multiple versions of these four datasets with the number of attributes ranging from 3 to 10 and their domain sizes ranging from $2^4$ to $2^{10}$.

\mypara{Competitors}
We compare HDG against TDG and all the baseline approaches including HIO, CALM, MSW and LHIO. In addition, we add a benchmark approach Uni which always outputs a uniform guess.
In particular, we set the branch factor $b=4$ for HIO and LHIO. For CALM, we choose to reconstruct high-dimensional marginals from 2-D ones.
\rv{To configure TDG and HDG with our guideline, we first set the recommended $\alpha_1=0.7$ and $\alpha_2 =0.03$. Then, for handling a dataset, we use its public information including the number of users $n$ and the number of attributes $d$ to obtain the $(n_i, m_i)$ ($i=1, 2$) according to the provided strategy in our guideline. Finally, given a privacy budget $\varepsilon$, we derive the values of $g_1$ and $g_2$ from the equations in our guideline.}
\rv{Note that except for HIO and Uni, all other approaches contain the consistency operation inside.}

\mypara{Utility Metric} \rv{We use the Mean Absolute Error (MAE) to measure the accuracy of estimated answers. Given a set $Q$ of range queries, it is computed as $\mbox{MAE} = \frac{1}{|Q|}\sum_{q\in Q}{|\eef_q- \rf_q|}$,}
\rv{where $\eef_q$ and $\rf_q$ are the estimated and true answers of query $q$, respectively.}

\mypara{Methodology}
To evaluate the performance of HDG, we randomly select a set $Q$ of $\lambda$-D range queries and calculate their MAE. We generate range queries with different dimensional query volumes denoted by $\omega$, which means the ratio of the specified interval to the domain size for each queried attribute.
In all subsequent experiments, unless explicitly stated, we use the following default values for other relevant parameters: $\varepsilon = 1.0$, $\omega=0.5$, $d = 6$, $c = 64$, $n= 10^6$, $\lambda = \rv{\{2, {4}\}}$ and $|Q|=200$.

We implemented all approaches using Python3.7. \rv{The source code of our approaches is publicly available at~\cite{technical_report_http}.}
All experiments were conducted on servers running Linux kernel version 5.0 with Intel Xeon Silver 4108 CPU @ 1.80GHz and 128GB memory. For each dataset and each approach, we repeat each experiment 10 times and report result mean and standard deviation. Note that standard deviation is invisible in most cases because the performance is stable in our results.
\rv{Besides, due to high MAEs, the results of HIO are automatically omitted in some figures for more noticeable differences among other approaches.}

\subsection{Overall Results}

Figure~\ref{fig:comparison_different_approaches_varying_epsilon} shows the results for comparing HDG against all the competitors under different $\varepsilon$ on all four datasets.
As expected, we can observe that except Uni, the accuracy of all other approaches becomes better (value of MAE gets lower) when $\varepsilon$ grows.  Among these approaches, HIO performs the worst, even worse than Uni in most of the cases.  On all datasets, our improved LHIO performs roughly one order of magnitude better than HIO in the low $\varepsilon$ region, but the improvement is less significant for a larger $\varepsilon$.  This is because when $\varepsilon$ is small, the consistency step of LHIO corrects many inconsistency; and when $\varepsilon$ gets larger, the error becomes less and so is the effect of the consistency step.
Moreover, CALM performs better than LHIO. The reason is that LHIO have much fewer users in each group than CALM under this setting, which incurs excessive noise canceling out the benefit of hierarchy.
We also observe that MSW can achieve a high accuracy on Bfive dataset  (Figure~\ref{fig:comparison_different_approaches_varying_epsilon}(c) and (d)), which indicates that the correlations among the attributes in Bfive dataset are weak. But we can see that the utility of HDG is still comparable to MSW, which confirms that HDG can also handle the datasets with low correlation well.

Figure~\ref{fig:comparison_different_approaches_varying_epsilon} shows that TDG and HDG have a clear advantage over other approaches; and HDG performs better than TDG.  Note that there are some jumping points of the two approaches.  This is because HDG and TDG choose different granularities based on $\varepsilon$ values and dataset sizes, and the choices, while generally good, are not optimal for every dataset at every $\varepsilon$ value.

\def\metric{MNAE}
\def\setwidth{0.47}
\def\setheight{0.36}
\def\mpagewidth{0.48}

\def\setwidth{\fwidth}
\def\setheight{\fheight}

\def\qdimlist{2, 4}
\def\dqv{0.5}
\def\dqvfile{5}
\def\qn{200}
\foreach \folder/\gran in {all_algorithm_vary_dqv}
	{
		\begin{figure*}[htb]
			\centering
			\foreach \datasetname/\capdatasetname in {ipums2018/Ipums, bfive/Bfive, normal/Normal, laplace/Laplace}
				{
					\foreach \atnum in {6}
						{
							\foreach \ds in {64}
								{
									\begin{minipage}[t]{\mpagewidth\linewidth}
										\centering
										\foreach \qdim in \qdimlist
										{
											\subfigure[\textbf{\capdatasetname}, \qdimstr= \qdim]
											{
												\includegraphics[width=\setwidth\columnwidth, height=\setheight\columnwidth]{\figurepath{\folder}/dn-\datasetname-un-1000000-an-\atnum-ds-\ds-fot-2-pf-2-qn-\qn-qd-\qdim-e-10-\metric.pdf}
											}
											\hspace{-0.1in}
										}
									\end{minipage}
								}
						}
				}
			\vspace{-0.1in}
			{\includegraphics[scale = 0.4]{\figurepath{\folder}/legend.pdf}}
			\caption{\rv{\textbf{Varying $\omega$ on all datasets under setting of \unstr = $10^6$,  \atnumstr = $6$, \dsstr = $64$, \epstr = $1.0$, \qdimstr = $2, 4$. MAEs are shown in log scale.}}}
			\label{fig:comparison_different_approaches_varying_dqv}
			\vspace{-4pt}
		\end{figure*}
	}

\def\metric{MNAE}
\def\setwidth{0.47}
\def\setheight{0.36}
\def\mpagewidth{0.48}

\def\setwidth{\fwidth}
\def\setheight{\fheight}

\def\qdimlist{2, 4}
\def\atnum{6}
\def\dqv{0.5}
\def\dqvfile{5}
\def\qn{200}
\foreach \folder/\gran in {all_algorithm_vary_domain_size}
	{
		\begin{figure*}[t]
			\centering
			\foreach \datasetname/\capdatasetname in {normal/Normal, laplace/Laplace}
				{
					\foreach \ds in {64}
						{
							\begin{minipage}[t]{\mpagewidth\linewidth}
								\centering
								\foreach \qdim in \qdimlist
								{
									\subfigure[\textbf{\capdatasetname}, \qdimstr= \qdim]
									{
										\includegraphics[width=\setwidth\columnwidth, height=\setheight\columnwidth]{\figurepath{\folder}/dn-\datasetname-un-1000000-an-\atnum-fot-2-pf-2-qn-\qn-dqv-\dqvfile-e-10-qd-\qdim-\metric.pdf}
									}
									\hspace{-0.1in}
								}
							\end{minipage}
						}
				}
			\vspace{-0.1in}
			{\includegraphics[scale = 0.4]{\figurepath{\folder}/legend.pdf}}
			\caption{\rv{\textbf{Varying $c$ on synthetic datasets under setting of \unstr = $10^6$,  \atnumstr = $6$,  \epstr = $1.0$, \dqvstr = $0.5$, \qdimstr = $2, 4$. MAEs are shown in log scale.}}}
			\label{fig:comparison_different_approaches_varying_domain_size}
			\vspace{-4pt}
		\end{figure*}
	}

\subsection{Impact of Different Parameters}

In this part, we compare different approaches under different parameter settings.
\rv{In general, these parameters including the dimensional query volume $\omega$, the domain size $c$ of an attribute, the number of attributes $d$, the query dimension $\lambda$ and the total number of users $n$ can also affect the performance of the approaches.}

\mypara{The impact of $\omega$}
Figure~\ref{fig:comparison_different_approaches_varying_dqv} shows the results varying $\omega$ from 0.1 to 0.9.
From Figure~\ref{fig:comparison_different_approaches_varying_dqv}, we can observe that HDG can consistently outperform all other approaches.
In general, for all approaches, their utilities degrade when $\omega$ increases. It is because there are more cells included in the range query and the noise error incurred by enforcing LDP grows. Moreover, we can observe that except for HIO, all LDP approaches have arch-like MAE trends, which means that their MAEs first increase and then decrease as $\omega$ increases.
This is due to the consistency operation, also observed in~\cite{DBLP:journals/corr/abs-1905-08320}.  In particular, when the queried area gets larger, with the enforcement of the consistency that the frequencies sum up to $1$, the result is essentially $1$ minus the un-queried areas.

\def\metric{MNAE}
\def\setwidth{0.47}
\def\setheight{0.36}
\def\mpagewidth{0.48}

\def\setwidth{\fwidth}
\def\setheight{\fheight}
\def\mpagewidth{\fmpagewidth}

\def\qdimlist{2, 4}
\def\dqv{0.5}
\def\dqvfile{5}	
\def\qn{100}
\foreach \folder/\gran in {all_algorithm_vary_attribute_num}
	{
		\begin{figure*}[htbp]
			\centering
			\foreach \datasetname/\capdatasetname in {ipums2018/Ipums, bfive/Bfive, normal/Normal, laplace/Laplace}
				{
					\foreach \ds in {64}
						{
							\begin{minipage}[t]{\mpagewidth\linewidth}
								\centering
								\foreach \qdim in \qdimlist
								{
									\subfigure[\textbf{\capdatasetname}, \qdimstr= \qdim]
									{
										\includegraphics[width=\setwidth\columnwidth, height=\setheight\columnwidth]{\figurepath{\folder}/dn-\datasetname-un-1000000-ds-\ds-fot-2-pf-2-qn-\qn-qd-\qdim-dqv-\dqvfile-e-10-\metric.pdf}
									}
									\hspace{-0.1in}
								}
							\end{minipage}
						}
				}
			\vspace{-0.1in}
			{\includegraphics[scale = 0.4]{\figurepath{\folder}/legend.pdf}}
			\caption{\rv{\textbf{Varying $d$ on all datasets under setting of \unstr = $10^6$,  \dsstr = $64$,  \epstr = $1.0$, \dqvstr = $0.5$, \qdimstr = $2, 4$. MAEs are shown in log scale.}}}
			\label{fig:comparison_different_approaches_varying_attribute_num}
			\vspace{-4pt}
		\end{figure*}
	}

\def\metric{MNAE}
\def\setwidth{0.47}
\def\setheight{0.36}
\def\mpagewidth{0.45}

\def\setwidth{\fwidth}
\def\setheight{\fheight}

\def\atnum{10}
\def\dqv{0.5}
\def\dqvfile{5}
\def\qn{200}
\foreach \folder in {all_algorithm_vary_query_dimension}
{
	\begin{figure*}[t]
		\centering
		\foreach \datasetname/\capdatasetname in {ipums2018/Ipums, bfive/Bfive, normal/Normal, laplace/Laplace}
		{
			\foreach \ds in {64}
			{
				\subfigure[\textbf{\capdatasetname}] 
				{
					\includegraphics[width=\setwidth\columnwidth, height=\setheight\columnwidth]{\figurepath{\folder}/dn-\datasetname-un-1000000-an-\atnum-ds-\ds-fot-2-pf-2-qn-\qn-dqv-\dqvfile-e-10-\metric.pdf}
				}
				\hspace{-0.1in} 
			}
		}
		\vspace{-0.1in}
		{\includegraphics[scale = 0.4]{\figurepath{\folder}/legend.pdf}}
		\caption{\rv{\textbf{Varying $\lambda$ on all datasets under setting of \unstr = $10^6$,  \atnumstr = $6$, \dsstr = $64$, \epstr = $1.0$, \dqvstr = $0.5$. MAEs are shown in log scale.}}}
		\label{fig:comparison_different_approaches_varying_query_dimension}
		\vspace{-4pt}
	\end{figure*}
}

\mypara{The impact of $c$}
Figure~\ref{fig:comparison_different_approaches_varying_domain_size} presents the results varying $c$ from $2^4$ to $2^{10}$ on synthetic datasets.
We can observe that HDG performs the best among all approaches. Moreover, the utility of HDG remains stable when $c$ becomes larger. It is because that the noise error and non-uniformity error do not change a lot for a grid as $c$ changes. As expected,  the MAEs of CALM and LHIO  become higher as $c$ increases, which is consistent with our analysis that more marginals included in the query lead to more LDP noise in the answer. We also find that MSW achieves higher utility when $c$ grows. That is because its advantage of  reporting values that are close to the true value with higher probabilities becomes more pronounced, especially for the Laplace dataset with spike distribution.

\mypara{The impact of $d$}
Figure~\ref{fig:comparison_different_approaches_varying_attribute_num} gives the results varying $d$ from 3 to 10. 
The relative order of different approaches are the same as we have already observed previously. We can observe that the MAEs of an LDP approach basically become higher when $d$ increases. The reason is that for a larger $d$, there are more user groups and fewer users in each group, which makes the amount of noise and sampling errors grow. In addition, we find an outlier at $d=10$ in Figure~\ref{fig:comparison_different_approaches_varying_attribute_num}(c) where the HDG's MAE at $d=10$ are smaller than those at $d=9$. This is due to the changes of the granularities. In particular, when $d$ increases from $9$ to $10$, the suggested granularities change from $(16, 4)$ to $(16, 2)$, which are more appropriate for Bfive dataset.

\mypara{The impact of $\lambda$}
Figure~\ref{fig:comparison_different_approaches_varying_query_dimension} studies the impact of $\lambda$ on the utility of each approach. 
We observe that the MAEs of LDP approaches decrease as $\lambda$ increases on real datasets (Figure~\ref{fig:comparison_different_approaches_varying_query_dimension}(a) and (b)). On synthetic datasets, the MAEs first grow and then drop along with the increment of  $\lambda$ (Figure~\ref{fig:comparison_different_approaches_varying_query_dimension}(c) and (d)). The reason can be explained as follows. Intuitively, when $\lambda$ becomes larger, there will be more estimation error included in the estimated answers. It is why the MAEs gradually grow at the beginning on synthetic datasets. However, for a relatively large $\lambda$, the true answer of a $\lambda$-D range query is close to zero. 
Due to the large amount of estimation error, the post-progress for removing negativity and inconsistency can also make the estimated answers approach zero and thus the MAEs are reduced. On real datasets, the effect of post-progress plays a decisive role since $\lambda=3$. We also find an outlier at $\lambda = 10$ in Figure~\ref{fig:comparison_different_approaches_varying_query_dimension}(a) where the HDG's MAE at $\lambda = 10$ are higher than that at $\lambda = 9$. 
The reason is that estimation error occurs when answering high dimensional range queries, and its feature that the magnitude is dependent on the dataset introduces uncertainty as mentioned in Section~\ref{subsec:theoretical_analysis}.

\noindent{\textbf{The impact of $n$.}}
Figure~\ref{fig:comparison_different_approaches_varying_user_num} shows the results varying $n$ from 100K to 10M on synthetic datasets.
Not surprisingly, for the approaches satisfying LDP, larger population can boost the accuracy of their results. 
We can observe that HDG consistently achieves the best performance among all approaches. It can be expected that when applying HDG to real-world applications where the population is large, we are able to achieve desirable performance.

\def\metric{MNAE}
\def\setwidth{0.47}
\def\setheight{0.36}
\def\mpagewidth{0.48}

\def\setwidth{\fwidth}
\def\setheight{\fheight}
\def\mpagewidth{\fmpagewidth}

\def\qdimlist{2, 4}
\def\atnum{6}
\def\dqv{0.5}
\def\dqvfile{5}
\def\qn{200}
\foreach \folder/\gran in {all_algorithm_vary_user_num}
{
	\begin{figure*}[htb]
		\centering
		\foreach \datasetname/\capdatasetname in {normal/Normal, laplace/Laplace}
		{
			\foreach \ds in {64}
			{
				\begin{minipage}[t]{\mpagewidth\linewidth}
					\centering
					\foreach \qdim in \qdimlist
					{
						\subfigure[\textbf{\capdatasetname}, \qdimstr= \qdim] 
						{
							\includegraphics[width=\setwidth\columnwidth, height=\setheight\columnwidth]{\figurepath{\folder}/dn-\datasetname-an-\atnum-ds-\ds-fot-2-pf-2-qn-\qn-dqv-\dqvfile-e-10-qd-\qdim-\metric.pdf}
						}
						\hspace{-0.1in}
					}
				\end{minipage}
			}
		}
		\vspace{-0.1in}
		{\includegraphics[scale = 0.4]{\figurepath{\folder}/legend.pdf}}
		\caption{\rv{\textbf{Varying $n$ on synthetic datasets under setting of \atnumstr = $6$, \dsstr = $64$, \epstr = $1.0$, \dqvstr = $0.5$, \qdimstr = $2, 4$. MAEs are shown in log scale.}}}
		\label{fig:comparison_different_approaches_varying_user_num}
		\vspace{-9pt}
	\end{figure*}
}

\def\metric{MNAE}
\def\setwidth{0.47}
\def\setheight{0.36}
\def\mpagewidth{0.45}

\def\setwidth{\fwidth}
\def\setheight{\fheight}

\def\atumlist{6}
\def\dslist{64}
\def\qdimlist{2}
\def\dqv{0.5}
\def\dqvfile{5}
\def\qn{200}
\foreach \folder in {grid_algorithm_vary_HG_1_2_way_granularity}
{
	\begin{figure*}[htpb]
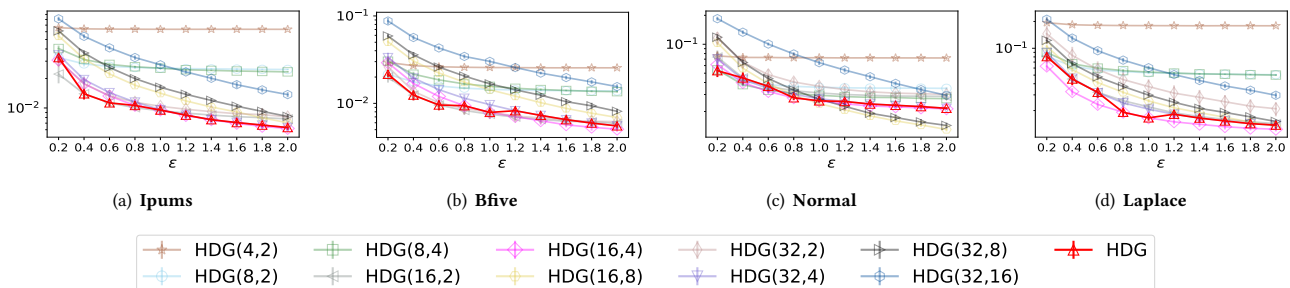

		\centering
		\foreach \datasetname/\capdatasetname in {ipums2018/Ipums, bfive/Bfive, normal/Normal, laplace/Laplace}
		{
			\foreach \atnum in {6}
			{
				\foreach \ds in {64}
				{
					\foreach \qdim in \qdimlist
					{
						\subfigure[\textbf{\capdatasetname}] 
						{
							\includegraphics[width=\setwidth\columnwidth, height=\setheight\columnwidth]{\figurepath{\folder}/dn-\datasetname-un-1000000-an-\atnum-ds-\ds-fot-2-pf-2-qn-\qn-qd-\qdim-dqv-\dqvfile-\metric.pdf}
						}
						\hspace{-0.22in}
					}
				}
			}
		}
		\vspace{-0.1in}
		{\includegraphics[scale = 0.4]{\figurepath{\folder}/legend.pdf}}
		\caption{\rv{\textbf{Verifying guideline in HDG under setting of \unstr = $10^6$, \atnumstr = $6$, \dsstr = $64$, \dqvstr = $0.5$, \qdimstr = $2$. MAEs are shown in log scale.}}}
		\label{fig:HDG_varying_granularity}
		\vspace{-9pt}
	\end{figure*}
}

\vspace{-10pt}
\subsection{Effectiveness of Guideline}
\label{subsec:exp_verify}

To evaluate the effectiveness of our proposed guideline for choosing granularities in HDG, we first enumerate all possible combinations of $g_1$ and  $g_2$ for a given domain size $c$.
Then, for each combination $(g_1 , g_2)$, we use it as the chosen granularities to implement a version of HDG, which is referred to as HDG($g_1 , g_2$). The approach labeled by HDG adopts granularities obtained from our proposed guideline under the suggested setting $\alpha_1=0.7$ and $\alpha_2=0.03$. Finally, we compare HDG with all the implemented versions to judge whether our guideline can provide good choices of granularities under different settings.

Figure~\ref{fig:HDG_varying_granularity} shows the results on 2-D range queries, which can avoid the the influence of estimation error.
From Figure~\ref{fig:HDG_varying_granularity}, we can see throughout the four datasets, HDG performs reasonably well for all $\varepsilon$ values. Although HDG may not perform best all the time, it can consistently achieve a very close accuracy to the best performing version, which confirms that our guideline can always give helpful guidance. We have also evaluated the effectiveness of our guideline under different $n, c$ and $d$; the results give similar conclusion, and are omitted due to space limitation.

\rv{We also conduct experiments for component-wise analysis to confirm the effectiveness of Phase 2 in HDG.
	Moreover, we experimentally study the performance of HDG under different parameter settings in our guideline to further verify the effectiveness of the recommended parameter settings. Furthermore, we investigate the convergence rates of Algorithms~\ref{alg:get_estimation_matrix} and~\ref{alg:weighted_update} to confirm their efficient convergence.
	In addition, we evaluate the performance of each approach on another two real datasets, namely Loan~\cite{data:loan_new} and Acs~\cite{data:acs}, and a new set of synthetic datasets varying the covariance between every two attributes ranging from 0 to 1 and confirm the superiority of HDG for handling diverse datasets. Due to space limitation, we show these experimental results and their analysis in the appendix.}

\vspace{-10pt}
\section{Related Work}
\label{sec:related}

Range queries have been widely studied under traditional DP~\cite{DBLP:conf/tcc/DworkMNS06}. Xiao et al.~\cite{DBLP:conf/icde/XiaoWG10} propose a framework Privelet, which employs wavelet transforms such as Haar wavelet to handle range queries.  Hay et al.~\cite{DBLP:journals/pvldb/HayRMS10} introduce the hierarchical intervals technique accompanied by constrained inference for ensuring consistency. Cormode et al.~\cite{DBLP:conf/icde/CormodePSSY12} utilize indexing methods such as quadtrees and kd-trees to generate spatial decompositions for describing the data distribution. Qardaji et al.~\cite{DBLP:journals/pvldb/QardajiYL13} provide a better understanding of using hierarchical methods for histogram publication.  
Li et al.~\cite{DBLP:journals/pvldb/LiHMW14} propose a two-stage approach DAWA utilizing a variant of the exponential mechanism to  partition the domain into uniform regions in the first stage, which cannot be done in LDP setting.
Qardaji et al.~\cite{DBLP:conf/icde/QardajiYL13} present an Adaptive Grids (AG) approach to release a synopsis for 2-D geospatial data and show that AG can perform better than those hierarchy approaches. Note that the idea of grid is also adopted in our approach HDG, but there are several differences between HDG and AG.
First, AG is only for 2-D data, and HDG is for multi-dimensional data, combining information from many 2-D grids.     
Moreover, for the first time, HDG proposes to combine information on both 1-D and 2-D grids to answer range queries. 
Finally, due to the feature of LDP setting, our HDG collects the information on grids by dividing users rather than the privacy budget and thus gives a novel analysis of different sources of errors and guideline.
\rv{For standardized evaluation of differential private algorithms that answering $1$-D and $2$-D range queries, Hay et al.~\cite{DBLP:conf/sigmod/HayMMCZ16} propose a novel evaluation framework DPBench. McKenna et al.~\cite{DBLP:journals/pvldb/McKennaMHM18} describe an algorithm HDMM, based on Matrix Mechanism~\cite{DBLP:conf/pods/LiHRMM10}, for answering workloads of predicate counting queries.} 

The notion of local differential privacy (LDP) was introduced in~\cite{DBLP:conf/focs/KasiviswanathanLNRS08}. Early works on LDP mainly focus on estimating frequencies of values of an attribute having a categorical domain~\cite{DBLP:conf/aistats/AcharyaSZ19,DBLP:conf/stoc/BassilyS15,DBLP:conf/ccs/ErlingssonPK14,DBLP:conf/uss/WangBLJ17,DBLP:journals/tit/YeB18}.
Wang et al.~\cite{DBLP:conf/uss/WangBLJ17} investigate these approaches and conclude that OLH is the state-of-the-art for a relatively large domain. More recently, for this problem, Wang et al.~\cite{DBLP:journals/pvldb/WangQDYHX20} propose a novel wheel mechanism, which has a same variance as OLH. For ordinal or numerical attributes, studies are mostly concentrated on mean estimation~\cite{DBLP:conf/focs/DuchiJW13,DBLP:conf/nips/DuchiWJ13,DBLP:conf/icde/WangXYZHSS019}. Only several works investigate range queries. For answering 1-D range queries on a singe attribute, Cormode et al.~\cite{DBLP:journals/pvldb/CormodeKS19} extend the ideas of hierarchical intervals and Haar wavelet transform to the LDP setting. Li et al.~\cite{DBLP:conf/sigmod/Li0LLS20} propose the Square Wave (SW) approach for reconstructing the distribution of an ordinal attribute. 
We have extended SW to answer multi-dimensional range queries in Section~\ref{subsec:multiplied_square_wave} and examined its performance. The most closely related work for answering multi-dimensional range queries is HIO proposed by Wang et al.~\cite{DBLP:conf/sigmod/WangDZHHLJ19}, which is designed for multi-dimensional analytical queries. We have considered HIO as a baseline approach and proposed an improvement of it in Section~\ref{sec:baseline_approaches}.

In addition, approaches~\cite{DBLP:conf/sigmod/CormodeKS18,DBLP:journals/tifs/RenYYYYMY18,DBLP:conf/ccs/ZhangWLHC18} for marginal release under LDP can be also used to answer multi-dimensional queries. Ren et al.~\cite{DBLP:journals/tifs/RenYYYYMY18} generalizes the Expectation Maximization algorithm for estimating joint distribution of two attributes. Cormode et al.~\cite{DBLP:conf/sigmod/CormodeKS18} refine and analyze how to release marginals via transformations under LDP. CALM proposed by Zhang et al.~\cite{DBLP:conf/ccs/ZhangWLHC18} is the state-of-art for marginal release under LDP. It adapts the ideas of consistency enforcement and maximum entropy estimation from PriView~\cite{DBLP:conf/sigmod/QardajiYL14} to LDP setting.
We have also analysed its performance in handling our problem in Section~\ref{subsec:calm}. 

LDP has been also applied to support other data analysis tasks, such as collecting frequent items or itemsets~\cite{DBLP:conf/nips/BassilyNST17,DBLP:conf/pods/BunNS18,DBLP:conf/icde/Gu2020protectionldp,DBLP:conf/icalp/HsuKR12,DBLP:conf/ccs/QinYYKXR16,DBLP:conf/icde/WangXYHSS018,DBLP:conf/infocom/0003HNWXY18,DBLP:conf/sp/WangLJ18}, locations~\cite{DBLP:conf/icde/ChenLQKJ16,DBLP:conf/cns/GuLC019}, key-value data~\cite{DBLP:conf/uss/Gu0C0020,DBLP:conf/sp/YeHMZ19}, social graphs~\cite{DBLP:conf/ccs/QinYYKX017,DBLP:conf/ccs/SunXKYQWY19}, 
\rv{linear query answers~\cite{DBLP:conf/aistats/Bassily19,DBLP:conf/stoc/EdmondsNU20,DBLP:journals/pvldb/McKennaMMM20}}, telemetry data~\cite{DBLP:conf/nips/DingKY17}, preference rankings~\cite{DBLP:conf/icde/YangCSCRL19} and evolving data~\cite{DBLP:conf/nips/JosephRUW18}. However, since they work on the problems that are different from ours, their approaches are not suitable for answering multi-dimensional range queries.

\section{Conclusions}
\label{sec:conc}

In this paper, we present TDG and HDG, two novel approaches for answering multi-dimensional range queries under LDP. We claim that TDG and HDG satisfy $\varepsilon$-LDP.  We theoretically analyse different sources of errors and provide a guideline for properly choosing granularities. Our results demonstrate the effectiveness of HDG.

\clearpage

\section*{Acknowledgement}
We sincerely thank the anonymous reviewers for their helpful comments and suggestions.
This work was supported by the National Natural
Science Foundation of China (Grant No. 61872045), the Innovation Research Group Project of NSFC (61921003), and the National Science Foundation (Grant No. 1640374 and No. 1931443).

{
	\bibliographystyle{styles/ACM-Reference-Format}
	\bibliography{ref_update_complete_concise_cof_arxiv}
}

\appendix

\def\metric{MNAE}

\def\setwidth{0.49}
\def\setheight{0.30}
\def\mpagewidth{0.48}

\def\atumlist{6}
\def\dslist{64}
\def\qdimlist{2, 4}
\def\dqv{0.5}
\def\dqvfile{5}
\def\qn{200}

\foreach \folder in {grid_algorithm_vary_noncons}
{
	\begin{figure*}[t]
		\centering
		\foreach \datasetname/\capdatasetname in {ipums2018/Ipums, bfive/Bfive, normal/Normal, laplace/Laplace}
		{
			\foreach \atnum in {6}
			{
				\foreach \ds in {64}
				{
					\begin{minipage}[t]{\mpagewidth\linewidth}
						\centering
						\foreach \qdim in \qdimlist
						{
							\subfigure[\textbf{\capdatasetname}, \qdimstr= \qdim] 
							{
								\includegraphics[width=\setwidth\columnwidth, height=\setheight\columnwidth]{\figurepath{\folder}/dn-\datasetname-un-1000000-an-\atnum-ds-\ds-fot-2-pf-2-qn-\qn-qd-\qdim-dqv-\dqvfile-\metric.pdf}
							}
							\hspace{-0.12in}
						}
					\end{minipage}
				}
			}
		}
		\vspace{-0.1in}
		{\includegraphics[scale = 0.4]{\figurepath{\folder}/legend.pdf}}
		\caption{\textbf{Component-wise analysis under setting of \unstr = $10^6$,  \atnumstr = $6$, \dsstr = $64$, \dqvstr = $0.5$, \qdimstr = $\qdimlist$. MAEs are shown in log scale.}}
		\label{fig:review_3_comparison_grid_approaches_varying_noncons}
	\end{figure*}
}

\section{Supplementary Analysis}

\subsection{Component-Wise Analysis}

There are two key components in our grid approaches TDG and HDG including Phase 2 (removing negativity and inconsistency) and Phase 3 (answering range queries). For component-wise analysis, we first remove Phase 2 of TDG and HDG to implement another two versions of TDG and HDG, which are referred to as Inconsistent TDG (ITDG) and Inconsistent HDG (IHDG), respectively. Then we compare TDG and HDG against ITDG and IHDG to evaluate the contribution of each component. Note that there may be negative outputs in ITDG and IHDG, which cannot guarantee the convergence of Algorithm~\ref{alg:get_estimation_matrix} (Building Response Matrix) or Algorithm~\ref{alg:weighted_update} (Answering $\lambda$-dimensional Range Query). Therefore, for ITDG and IHDG, we set the the maximum number of iterations as 100 in Algorithms~\ref{alg:get_estimation_matrix} and~\ref{alg:weighted_update} in our experiments.

Figure~\ref{fig:review_3_comparison_grid_approaches_varying_noncons} shows the results varying $\varepsilon$ from $0.2$ to $2.0$. From Figure~\ref{fig:review_3_comparison_grid_approaches_varying_noncons}, we observe that ITDG and TDG achieve nearly the same accuracy in all cases. This is because in ITDG and TDG,  the grids are coarse-grained and each user group has more population, which produce very few negative outputs, canceling out the benefit of Phase 2. We also see that the performance of IHDG is unstable. The reason is that the negative outputs in IHDG have a great influence on the process of Weighted Update in Algorithms~\ref{alg:get_estimation_matrix} and~\ref{alg:weighted_update}. Even a very small negative value can change the convergence trend of the whole iteration process. 
In most cases, HDG can achieve a clearly better and more stable accuracy than IHDG, which confirms the effectiveness of Phase 2. Besides, we find some exceptions where IHDG has a lower error than HDG in Figures~\ref{fig:review_3_comparison_grid_approaches_varying_noncons} (e-h). The reason can be explained as follows. As described in Section 4.5, the Weighted Update process in Algorithm 1 and 2 may introduce uncertainty on accuracy due to its dependency on the distribution of dataset. We have also performed the component-wise analysis when $\lambda= 3, 5, 6$ and the results give similar conclusion. Due to space limitation, we omit them here.

\subsection{Standard Error Distribution of TDG and HDG}

\def\metric{MNAE}
\def\setwidth{0.49}
\def\setheight{0.30}
\def\mpagewidth{0.48}

\def\qdimlist{2, 4}
\def\dqv{0.5}
\def\dqvfile{5}
\def\qn{200}
\foreach \folder/\gran in {grid_algorithm_std_error_distribution}
{
	\begin{figure*}[ht]
		\centering
		\foreach \datasetname/\capdatasetname in {ipums2018/Ipums, bfive/Bfive, normal/Normal, laplace/Laplace}
		{
			\foreach \atnum in {6}
			{
				\foreach \ds in {64}
				{
					\begin{minipage}[t]{\mpagewidth\linewidth}
						\centering
						\foreach \qdim in \qdimlist
						{
							\subfigure[\textbf{\capdatasetname}, \qdimstr= \qdim] 
							{
								\includegraphics[width=\setwidth\columnwidth, height=\setheight\columnwidth]{\figurepath{\folder}/an2-Grid_2_opt-dn-\datasetname-un-1000000-an-\atnum-ds-\ds-fot-2-e-10-gan-2-pf-2-qn-\qn-qd-\qdim-dqv-5-\metric.pdf}
							}
							\hspace{-0.12in}
						}
					\end{minipage}
					\hspace{-0.15in}
				}
			}
		}
		\caption{\textbf{TDG standard error distribution under setting of \unstr = $10^6$,  \atnumstr = $6$, \dsstr = $64$, \epstr = $1.0$, \dqvstr = $0.5$, \qdimstr = $\qdimlist$.}}
		\label{fig:TDG_std_error_distribution}
	\end{figure*}
}

For a single query $q$ in the given query set $Q$, we calculate its standard (absolute) error as $|\eef_q- \rf_q|$, where $\eef_q$ and $\rf_q$ are the estimated and true answers of query $q$, respectively. 
To evaluate the distribution of the standard errors, we also run each experiment 10 times. For each query $q$, we consider the mean of 10 standard errors as the final one. 
Figures~\ref{fig:TDG_std_error_distribution} and \ref{fig:HDG_std_error_distribution} show the distribution of the standard errors of TDG and HDG, respectively. 


\def\metric{MNAE}
\def\metric{MNAE}
\def\setwidth{0.49}
\def\setheight{0.30}
\def\mpagewidth{0.48}

\def\qdimlist{2, 4}
\def\dqv{0.5}
\def\dqvfile{5}
\def\qn{200}
\foreach \folder/\gran in {grid_algorithm_std_error_distribution}
{
	\begin{figure*}[ht]
		\centering
		\foreach \datasetname/\capdatasetname in {ipums2018/Ipums, bfive/Bfive, normal/Normal, laplace/Laplace}
		{
			\foreach \atnum in {6}
			{
				\foreach \ds in {64}
				{
					\begin{minipage}[t]{\mpagewidth\linewidth}
						\centering
						\foreach \qdim in \qdimlist
						{
							\subfigure[\textbf{\capdatasetname}, \qdimstr= \qdim] 
							{
								\includegraphics[width=\setwidth\columnwidth, height=\setheight\columnwidth]{\figurepath{\folder}/an2-Grid_1+2_opt-dn-\datasetname-un-1000000-an-\atnum-ds-\ds-fot-2-e-10-gan-2-pf-2-qn-\qn-qd-\qdim-dqv-5-\metric.pdf}
							}
							\hspace{-0.12in}
						}
					\end{minipage}
					\hspace{-0.15in}
				}
			}
		}
		\caption{\textbf{HDG standard error distribution under setting of \unstr = $10^6$,  \atnumstr = $6$, \dsstr = $64$, \epstr = $1.0$, \dqvstr = $0.5$, \qdimstr = $\qdimlist$.}}
		\label{fig:HDG_std_error_distribution}
	\end{figure*}
}

\subsection{Answering Full 2-D Marginals and Range Queries}

\def\metric{MNAE}
\def\setwidth{0.47}
\def\setheight{0.36}

\def\setwidth{0.49}
\def\setheight{0.30}
\def\mpagewidth{0.48}

\def\atumlist{6}
\def\dslist{64}
\def\qdimlist{2}
\def\dqv{0.5}
\def\dqvfile{0}
\def\qn{10000}
\foreach \folder in {all_algorithm_vary_epsilon}
{
	\begin{figure*}[t]
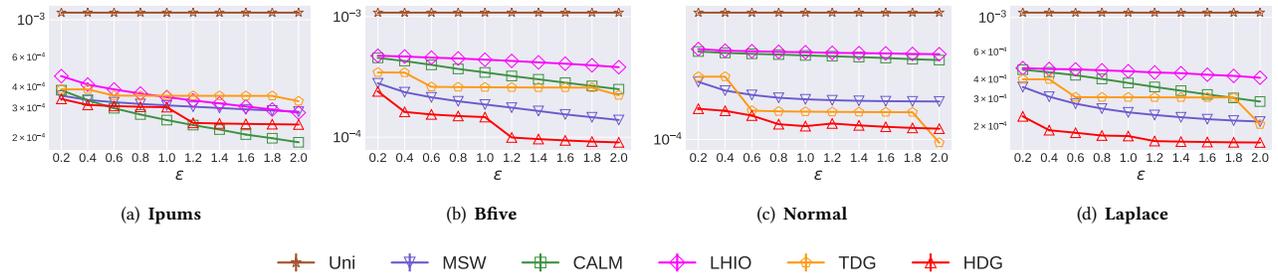

		\centering
		\foreach \datasetname/\capdatasetname in {ipums2018/Ipums, bfive/Bfive, normal/Normal, laplace/Laplace}
		{
			\foreach \atnum in {6}
			{
				\foreach \ds in {64}
				{
					\foreach \qdim in \qdimlist
					{
						\subfigure[\textbf{\capdatasetname}] 
						{
							\includegraphics[width=\setwidth\columnwidth, height=\setheight\columnwidth]{\figurepath{\folder}/dn-\datasetname-un-1000000-an-\atnum-ds-\ds-fot-2-pf-2-qn-\qn-qd-\qdim-dqv-\dqvfile-\metric.pdf}
						}
						\hspace{-0.3in}
					}
				}
			}
		}
		\vspace{-0.1in}
		{\includegraphics[scale = 0.4]{\figurepath{\folder}/legend-no_PHDR.pdf}}
		\caption{\textbf{Full 2-D marginal queries comparison under setting of \unstr = $10^6$, \atnumstr = $6$, \dsstr = $64$, \qdimstr = $2$. MAEs are shown in log scale.}}
		\label{fig:review_3_full_2D_marginals_comparison}
		\vspace{12pt}
	\end{figure*}
}

\def\metric{MNAE}
\def\setwidth{0.47}
\def\setheight{0.36}

\def\setwidth{0.49}
\def\setheight{0.30}
\def\mpagewidth{0.48}

\def\atumlist{6}
\def\dslist{64}
\def\qdimlist{2}
\def\dqv{0.5}
\def\dqvfile{5}
\def\qn{10000}
\foreach \folder in {all_algorithm_vary_epsilon}
{
	\begin{figure*}[t]
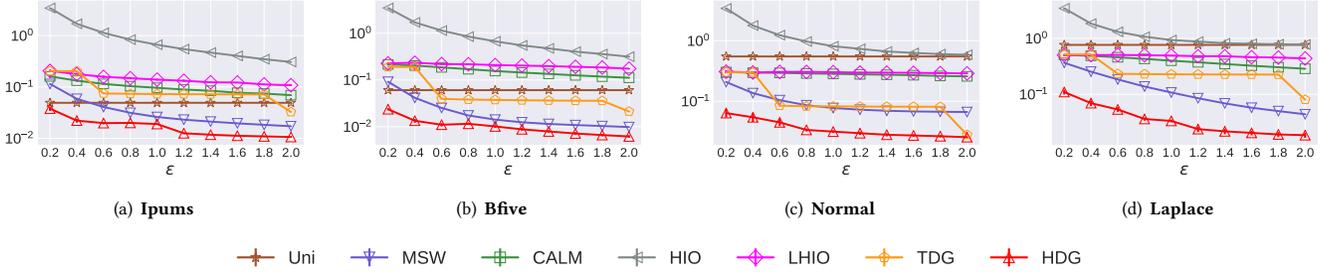

		\centering
		\foreach \datasetname/\capdatasetname in {ipums2018/Ipums, bfive/Bfive, normal/Normal, laplace/Laplace}
		{
			\foreach \atnum in {6}
			{
				\foreach \ds in {64}
				{
					\foreach \qdim in \qdimlist
					{
						\subfigure[\textbf{\capdatasetname}] 
						{
							\includegraphics[width=\setwidth\columnwidth, height=\setheight\columnwidth]{\figurepath{\folder}/dn-\datasetname-un-1000000-an-\atnum-ds-\ds-fot-2-pf-2-qn-\qn-qd-\qdim-dqv-\dqvfile-\metric.pdf}
						}
						\hspace{-0.2in} 
					}
				}
			}
		}
		\vspace{-0.1in}
		{\includegraphics[scale = 0.4]{\figurepath{\folder}/legend.pdf}}
		\caption{\textbf{Full 2-D range queries comparison under setting of \unstr = $10^6$, \atnumstr = $6$, \dsstr = $64$, \dqvstr = $0.5$, \qdimstr = $2$. MAEs are shown in log scale.}}
		\label{fig:review_3_full_2D_range_queries_comparison}
	\end{figure*}
}

\def\metric{MNAE}
\def\setwidth{0.47}
\def\setheight{0.36}
\def\mpagewidth{0.45}

\def\setwidth{0.49}
\def\setheight{0.30}
\def\mpagewidth{0.48}

\def\atnum{10}
\def\dqv{0.5}
\def\dqvfile{3}
\def\qn{150}
\foreach \folder in {all_algorithm_vary_query_dimension}
{
	\begin{figure*}[t]
		\centering
		\foreach \datasetname/\capdatasetname in {ipums2018/Ipums, bfive/Bfive, normal/Normal, laplace/Laplace}
		{
			\foreach \ds in {64}
			{
				\subfigure[\textbf{\capdatasetname}] 
				{
					\includegraphics[width=\setwidth\columnwidth, height=\setheight\columnwidth]{\figurepath{\folder}/dn-\datasetname-un-1000000-an-\atnum-ds-\ds-fot-2-pf-2-qn-\qn-dqv-\dqvfile-e-10-\metric.pdf}
				}
				\hspace{-0.1in} 
			}
		}
		\vspace{-0.1in}
		{\includegraphics[scale = 0.4]{\figurepath{\folder}/legend.pdf}}
		\caption{\textbf{0-count queries comparison under setting of \unstr = $10^6$,  \atnumstr = $6$, \dsstr = $64$, \epstr = $1.0$, \dqvstr = $0.3$. MAEs are shown in log scale.}}
		\label{fig:review_3_0_count_comparison_different_approaches_varying_query_dimension}
	\end{figure*}
}

\def\metric{MNAE}
\def\setwidth{0.47}
\def\setheight{0.36}
\def\mpagewidth{0.45}

\def\setwidth{0.49}
\def\setheight{0.30}
\def\mpagewidth{0.48}

\def\atnum{10}
\def\dqv{0.5}
\def\dqvfile{7}
\def\qn{150}
\foreach \folder in {all_algorithm_vary_query_dimension}
{
	\begin{figure*}[t]
		\centering
		\foreach \datasetname/\capdatasetname in {ipums2018/Ipums, bfive/Bfive, normal/Normal, laplace/Laplace}
		{
			\foreach \ds in {64}
			{
				\subfigure[\textbf{\capdatasetname}] 
				{
					\includegraphics[width=\setwidth\columnwidth, height=\setheight\columnwidth]{\figurepath{\folder}/dn-\datasetname-un-1000000-an-\atnum-ds-\ds-fot-2-pf-2-qn-\qn-dqv-\dqvfile-e-10-\metric.pdf}
				}
				\hspace{-0.1in} 
			}
		}
		\vspace{-0.1in}
		{\includegraphics[scale = 0.4]{\figurepath{\folder}/legend.pdf}}
		\caption{\textbf{Non-0-count queries comparison under setting of \unstr = $10^6$,  \atnumstr = $6$, \dsstr = $64$, \epstr = $1.0$, \dqvstr = $0.7$. MAEs are shown in log scale.}}
		\label{fig:review_3_non_0_count_comparison_different_approaches_varying_query_dimension}
	\end{figure*}
}

To evaluate the performance of HDG for answering full 2-D marginal queries, we fix \unstr = $10^6$, \atnumstr = $6$, \dsstr = $64$, \qdimstr = $2$, and generate all $\binom{d}{2} \cdot c^2 =  \binom{6}{2} \cdot 64^2 = 61440$ 2-D marginal queries. Figure~\ref{fig:review_3_full_2D_marginals_comparison} shows the results of full 2-D marginal queries varying $\varepsilon$ from 0.2 to 1.0. 
The results of HIO are omitted due to its high errors (larger than $10^{-2}$).
From Figure~\ref{fig:review_3_full_2D_marginals_comparison}, we observe that CALM can perform a little better than HDG on Ipums dataset. However, on other three datasets Bfive, Normal and Laplace, even TDG achieves higher accuracy than CLAM, and HDG still performs best among all approaches. The reason can be explained as follows. 
As the state-of-the-art for marginal release under LDP, CALM directly employs OLH to collect each 2-D marginal. Therefore, the expected squared error of each collected marginal is dominated by the variance of OLH (say $Var_0$). For TDG, since the 2-D domain of each attribute pair is partitioned into some cells, the expected squared error of each cell's frequency is also dominated by the variance of OLH, which equals that of each marginal in CALM, i.e., $Var_0$. For a cell which contains $\gamma$ marginals, every marginal within it has two main sources of errors. One is non-uniformity error due to the assumption that the values in a cell are uniformly distributed. The more uniformly distributed the values in each cell are, the smaller this error is. The other one is noise error due to the usage of OLH. Apparently, the sum of squared noise error of each marginal in the cell equals the squared error of its frequency $Var_0$. Under the uniformity assumption, the squared noise error of each marginal can be considered as $\frac{Var_0}{\gamma}$, which is smaller than $Var_0$ in CALM. Back to our results, that TDG performs better than CALM on datasets Bfive, Normal and Laplace is because these three datasets have more uniform distribution in each cell, leading to small non-uniformity error. By introducing 1-D grids, HDG further reduces the non-uniformity error and thus achieves much better accuracy than TDG. On Ipums dataset, while HDG cannot beat CALM, the utility of HDG is still comparable to CALM.

Besides, we evaluate the performance of HDG for answering full 2-D range queries. In particular, we fix \unstr = $10^6$, \atnumstr = $6$, \dsstr = $64$, \dqvstr = $0.5$, \qdimstr = $2$ and generate all $\binom{d}{2} \cdot (c \cdot \omega)^2 =  \binom{6}{2} \cdot 32^2 = 15360$ 2-D range queries.
Figure~\ref{fig:review_3_full_2D_range_queries_comparison} shows the results of full 2-D range queries of \dqvstr = $0.5$ varying $\varepsilon$ from 0.2 to 1.0. As expected, HDG can achieve the best performance, which confirms the superiority of HDG for answering range queries.

\def\metric{MNAE}
\def\setwidth{0.47}
\def\setheight{0.36}
\def\mpagewidth{0.45}

\def\setwidth{0.49}
\def\setheight{0.30}
\def\mpagewidth{0.48}

\def\atumlist{6}
\def\dslist{64}
\def\qdimlist{2}
\def\dqv{0.5}
\def\dqvfile{5}
\def\qn{200}
\foreach \folder in {grid_algorithm_vary_HG_sigma}
{
	\begin{figure*}[t]
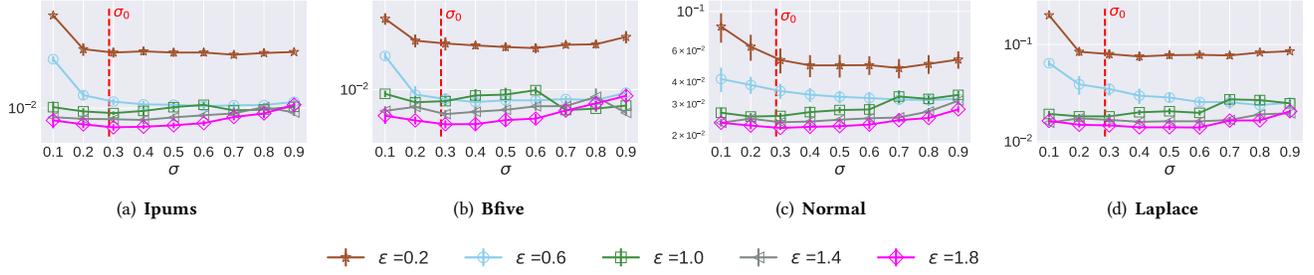

		\centering
		\foreach \datasetname/\capdatasetname in {ipums2018/Ipums, bfive/Bfive, normal/Normal, laplace/Laplace}
		{
			\foreach \atnum in {6}
			{
				\foreach \ds in {64}
				{
					\foreach \qdim in \qdimlist
					{
						\subfigure[\textbf{\capdatasetname}] 
						{
							\includegraphics[width=\setwidth\columnwidth, height=\setheight\columnwidth]{\figurepath{\folder}/dn-\datasetname-un-1000000-an-\atnum-ds-\ds-fot-2-pf-2-qn-\qn-qd-\qdim-dqv-\dqvfile-\metric.pdf}
						}
						\hspace{-0.24in}
					}
				}
			}
		}
		\vspace{-0.1in}
		{\includegraphics[scale = 0.4]{\figurepath{\folder}/legend.pdf}}
		\caption{\textbf{Justifying $n_1$ and $n_2$ in HDG under setting of \unstr = $10^6$, \atnumstr = $6$, \dsstr = $64$, \dqvstr = $0.5$, \qdimstr = $2$. MAEs are shown in log scale.}}
		\label{fig:review_3_justifying_choice_of_n}
	\end{figure*}
}

\def\metric{MNAE}
\def\setwidth{0.47}
\def\setheight{0.36}

\def\setwidth{0.23}
\def\setheight{0.14}
\def\mpagewidth{1.0}

\def\qdim{2}
\def\dqv{0.5}
\def\dqvfile{5}
\def\qn{200}
\foreach \folder/\gran in {grid_algorithm_vary_HG_1_2_way_granularity}
{
	\begin{figure*}[htb]
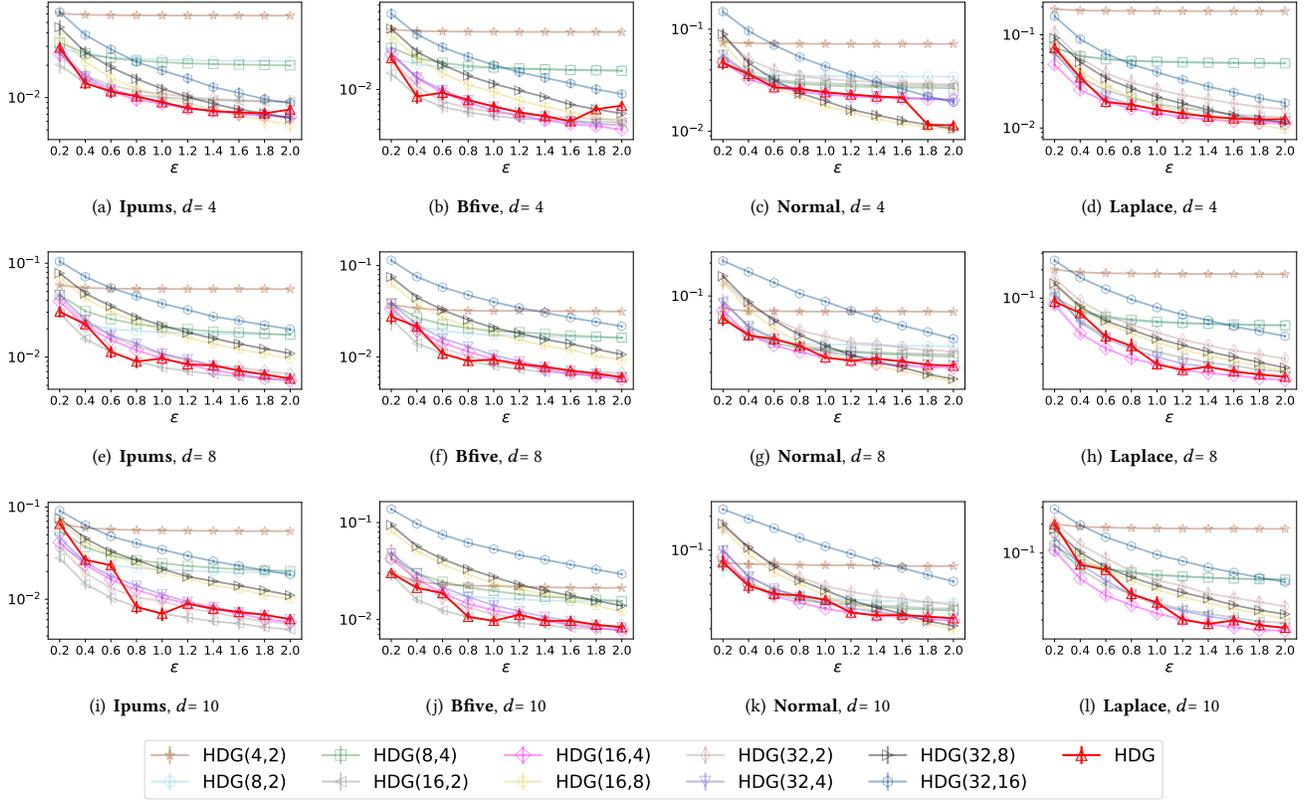

		\centering
		\foreach \atnum in {4, 8, 10}
		{
			\foreach \ds in {64}
			{
				\begin{minipage}[t]{\mpagewidth\linewidth}
					\centering
					\foreach \datasetname/\capdatasetname in {ipums2018/Ipums, bfive/Bfive, normal/Normal, laplace/Laplace}
					{
						\subfigure[\textbf{\capdatasetname}, \atnumstr= \atnum] 
						{
							\includegraphics[width=\setwidth\columnwidth, height=\setheight\columnwidth]{\figurepath{\folder}/dn-\datasetname-un-1000000-an-\atnum-ds-\ds-fot-2-pf-2-qn-\qn-qd-2-dqv-\dqvfile-\metric.pdf}
						}
					}
				\end{minipage}
			}
		}
		\vspace{-0.1in}
		{\includegraphics[scale = 0.4]{\figurepath{\folder}/legend.pdf}}
		\caption{\textbf{Verifying guideline in HDG under setting of \unstr = $10^6$, \atnumstr = $4, 8, 10$, \dsstr = $64$, \dqvstr = $0.5$, \qdimstr = $2$. MAEs are shown in log scale.}}
		\label{fig:review_3_HDG_varying_granularity}
	\end{figure*}
}

\subsection{Answering 0-Count and Non-0-Count High Dimensional Queries}

To evaluate the performance of HDG for answering high dimensional queries that have smaller counts and higher counts, respectively, we randomly select sets of 0-count range queries of $\omega = 0.3$ and sets of non-0-count range queries of $\omega = 0.7$. Figures~\ref{fig:review_3_0_count_comparison_different_approaches_varying_query_dimension} and~\ref{fig:review_3_non_0_count_comparison_different_approaches_varying_query_dimension} show the results on 0-count queries and non-0-count queries, respectively, varying $\lambda$ from 6 to 10. The results of HIO are omitted due to its high errors (larger than 1).
From Figure~\ref{fig:review_3_0_count_comparison_different_approaches_varying_query_dimension}, we observe that compared with the baseline approaches, the advantage of HDG is not obvious, since all approaches can achieve very low errors (less than $10^{-4}$). It is because the removing negativity and inconsistency operation help make the estimated answers approach zero. 
From Figure~\ref{fig:review_3_non_0_count_comparison_different_approaches_varying_query_dimension}, we can see that HDG typically obtain better results than the existing approaches.
In particular, from Figure~\ref{fig:review_3_non_0_count_comparison_different_approaches_varying_query_dimension}(a) and (b), we observe that the MAEs tend to be smaller as $\lambda$ grows, while the opposite trend is found in Figure~\ref{fig:review_3_non_0_count_comparison_different_approaches_varying_query_dimension}(c) and (d). The reason can be explained as follows. For synthetic datasets Normal and Laplace, when $\lambda$ becomes larger, there will be more estimation errors included in the estimated answers. It is why the MAEs of TDG and HDG gradually grow. However, for real datasets Ipums and Bfive, the overall real answers of queries become much smaller when $\lambda$ is larger. This makes the effect of the removing negativity and inconsistency operation more pronounced, and thus the MAEs tend to be smaller.

\def\metric{MNAE}
\def\setwidth{0.47}
\def\setheight{0.36}
\def\mpagewidth{0.45}

\def\setwidth{0.49}
\def\setheight{0.30}
\def\mpagewidth{0.48}

\def\atumlist{6}
\def\dslist{64}
\def\qdimlist{2}
\def\dqv{0.5}
\def\dqvfile{5}
\def\qn{200}
\foreach \folder in {HDG_converge_rate_response_matrix}
{
	\begin{figure*}[t]
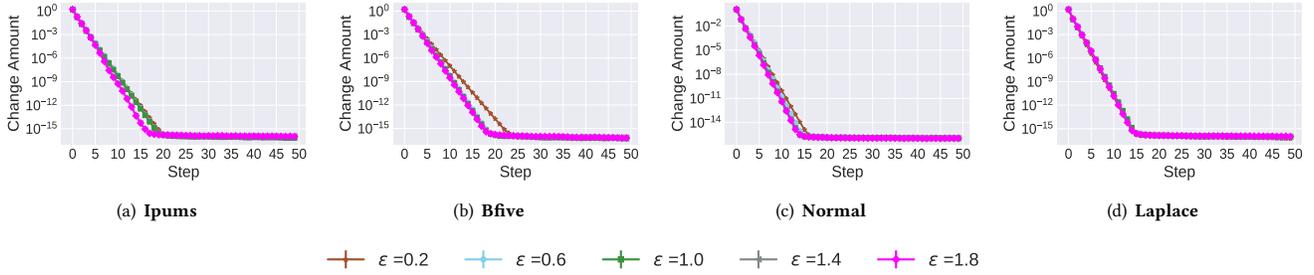

		\centering
		\foreach \datasetname/\capdatasetname in {ipums2018/Ipums, bfive/Bfive, normal/Normal, laplace/Laplace}
		{
			\foreach \atnum in {6}
			{
				\foreach \ds in {64}
				{
					\foreach \qdim in \qdimlist
					{
						\subfigure[\textbf{\capdatasetname}] 
						{
							\includegraphics[width=\setwidth\columnwidth, height=\setheight\columnwidth]{\figurepath{\folder}/dn-\datasetname-un-1000000-an-\atnum-ds-\ds-fot-2-pf-2-\metric.pdf}
						}
						\hspace{-0.24in}
					}
				}
			}
		}
		\vspace{-0.1in}
		{\includegraphics[scale = 0.4]{\figurepath{\folder}/legend.pdf}}
		\caption{\textbf{Convergence rate of Algorithm 1 under setting of \unstr = $10^6$, \atnumstr = $6$, \dsstr = $64$. Results are shown in log scale.}}
		\label{fig:review_3_converge_rate_of_algorithm_1}
	\end{figure*}
}

\def\atumlist{6}
\def\dslist{64}
\def\qdimlist{4}
\def\dqv{0.5}
\def\dqvfile{5}
\def\qn{200}
\foreach \folder in {HDG_converge_rate_answer_query}
{
	\begin{figure*}[t]
		\centering
		\foreach \datasetname/\capdatasetname in {ipums2018/Ipums, bfive/Bfive, normal/Normal, laplace/Laplace}
		{
			\foreach \atnum in {6}
			{
				\foreach \ds in {64}
				{
					\foreach \qdim in \qdimlist
					{
						\subfigure[\textbf{\capdatasetname}] 
						{
							\includegraphics[width=\setwidth\columnwidth, height=\setheight\columnwidth]{\figurepath{\folder}/dn-\datasetname-un-1000000-an-\atnum-ds-\ds-fot-2-pf-2-qn-\qn-qd-\qdim-dqv-\dqvfile-\metric.pdf}
						}
						\hspace{-0.24in}
					}
				}
			}
		}
		\vspace{-0.1in}
		{\includegraphics[scale = 0.4]{\figurepath{\folder}/legend.pdf}}
		\caption{\textbf{Convergence rate of Algorithm 2 under setting of \unstr = $10^6$, \atnumstr = $6$, \dsstr = $64$, \qdimstr = $\qdimlist$. Results are shown in log scale.}}
		\label{fig:review_3_converge_rate_of_algorithm_2}
	\end{figure*}
}

\subsection{Justifying Parameter Choices in Guideline}
Firstly, we justify the choices of $n_1$ and $n_2$ in the guideline.
For $d$ attributes and $n$ users, HDG constructs $d$ 1-D grids and $\binom{d}{2}$ 2-D grids, corresponding to $m_1$ and $m_2$ user groups, respectively. There are $n_1$ ($n_2$) users in $m_1$ ($m_2$) user groups, where $n_1 + n_2 = n$. As mentioned in Section 4.6, for simplicity, we make each user group have the same population, i.e., $\frac{{{n_1}}}{{{m_1}}} = \frac{{{n_2}}}{{{m_2}}} = \frac{n}{{d + \binom{d}{2}}}$ for HDG, as the default setting. 

\vspace{3pt}
To evaluate the effectiveness of this setting, we introduce a new variable $\sigma$, which denotes the proportion of $n_1$ to $n$, i.e., $\frac{{{n_1}}}{n} = \sigma$. Note that in the default setting, the value of $\sigma$ equals $\sigma_0 = \frac{d}{{d + \binom{d}{2}}}$.
Figure~\ref{fig:review_3_justifying_choice_of_n} shows the results of HDG with different $\varepsilon$  varying $\sigma$ from $0.1$ to $0.9$. From Figure~\ref{fig:review_3_justifying_choice_of_n}, we can see that in all cases, the values of $\sigma$ ranging from 0.2 to 0.6 can make HDG consistently achieve nearly the best performance, which confirms the effectiveness of our default setting ($\sigma_0 = 0.2857$ under this experimental setting).

Second, we verify the effectiveness of setting of $\alpha_1$ and $\alpha_2$. We have to point out that it is infeasible to find an identical setting of $\alpha_1$ and $\alpha_2$ which is optimal for all datasets due to their different distribution. For obtaining appropriate values of $\{\alpha_1, \alpha_2\}$, we tuned them on synthetic datasets under different setting of $n, c, d$ and recommended that $\alpha_1=0.7$ and $\alpha_2 =0.03$ can typically achieve good performance. Note that this process does not leak any real users' private information, since it is only run on synthetic datasets. To verify the effectiveness of the recommended setting, we further run the set of experiments in Section 5.4 under setting of $d = 4, 8, 10$ and present the results in Figure~\ref{fig:review_3_HDG_varying_granularity}. We can observe that HDG consistently achieve a very close accuracy to the best performing version, which confirms the effectiveness of our recommended values of $\alpha_1$ and $\alpha_2$.

\subsection{Convergence of Algorithms 1 and 2}
\label{appendix:convergence_algorithms}


The estimation processes in Algorithms 1 (Building Response Matrix) and 2 (Answering $\lambda$-Dimensional Range Query) correspond to linear programming problems, which include an under-specified equation system. As an efficient method for solving such problems, Weighted Update method has been proved to converge in~\cite{DBLP:conf/nips/HardtLM12}.
The intuition behind this method is to start from a uniform distribution of all variables and prune the distribution to satisfy the equations.

To evaluate the convergence rate of Algorithm 1 in HDG, we first calculate the sum of the changes of all elements in a response matrix, i.e., change amount, after each iteration step. For each step, we use the average of its corresponding change amount of all $\binom{d}{2}$ response matrices as the reported change amount. 
Figure~\ref{fig:review_3_converge_rate_of_algorithm_1} shows the convergence rate of Algorithm 1. From Figure~\ref{fig:review_3_converge_rate_of_algorithm_1}, we can observe that Algorithm 1 converges after twenty steps, which confirms the efficiency of Weighted Update method.

To evaluate the convergence rate of Algorithm 2 in HDG, we first calculate change amount of a range query's answer vector after each iteration step.
For each step, we use the average of its corresponding change amount of all queries as the reported change amount. 
Figure~\ref{fig:review_3_converge_rate_of_algorithm_2} shows the convergence rate of Algorithm 2 when $\lambda = 4$. We can see that the change rate of the change amount becomes very slow after twenty steps. Besides, we also run experiments to evaluate the convergence rate of Algorithm 2 when $\lambda = 3, 5, 6$. The results give similar conclusion and are omitted.

\subsection{Results on New Real Datasets}

\def\metric{MNAE}

\def\setwidth{0.49}
\def\setheight{0.30}
\def\mpagewidth{0.48}

\def\atumlist{6}
\def\dslist{64}
\def\qdimlist{2, 4}
\def\dqv{0.5}
\def\dqvfile{5}
\def\qn{200}

\foreach \folder in {all_algorithm_vary_epsilon}
{
	\begin{figure*}[t]
		\centering
		\foreach \datasetname/\capdatasetname in {loan/Loan, acs15/Acs}
		{
			\foreach \atnum in {6}
			{
				\foreach \ds in {64}
				{
					\begin{minipage}[t]{\mpagewidth\linewidth}
						\centering
						\foreach \qdim in \qdimlist
						{
							\subfigure[\textbf{\capdatasetname}, \qdimstr= \qdim] 
							{
								\includegraphics[width=\setwidth\columnwidth, height=\setheight\columnwidth]{\figurepath{\folder}/dn-\datasetname-un-1000000-an-\atnum-ds-\ds-fot-2-pf-2-qn-\qn-qd-\qdim-dqv-\dqvfile-\metric.pdf}
							}
							\hspace{-0.12in}
						}
					\end{minipage}
				}
			}
		}
		\vspace{-0.1in}
		{\includegraphics[scale = 0.4]{\figurepath{\folder}/legend.pdf}}
		\caption{\textbf{Varying $\varepsilon$ on new real datasets under setting of \unstr = $10^6$,  \atnumstr = $6$, \dsstr = $64$, \dqvstr = $0.5$, \qdimstr = $\qdimlist$. MAEs are shown in log scale.}}
		\label{fig:review_2_comparison_different_approaches_varying_epsilon}
	\end{figure*}
}

\def\setwidth{0.49}
\def\setheight{0.30}
\def\mpagewidth{0.48}

\def\qdimlist{2, 4}
\def\dqv{0.5}
\def\dqvfile{5}
\def\qn{200}
\foreach \folder/\gran in {all_algorithm_vary_dqv}
{
	\begin{figure*}[t]
		\centering
		\foreach \datasetname/\capdatasetname in {loan/Loan, acs15/Acs}
		{
			\foreach \atnum in {6}
			{
				\foreach \ds in {64}
				{
					\begin{minipage}[t]{\mpagewidth\linewidth}
						\centering
						\foreach \qdim in \qdimlist
						{
							\subfigure[\textbf{\capdatasetname}, \qdimstr= \qdim] 
							{
								\includegraphics[width=\setwidth\columnwidth, height=\setheight\columnwidth]{\figurepath{\folder}/dn-\datasetname-un-1000000-an-\atnum-ds-\ds-fot-2-pf-2-qn-\qn-qd-\qdim-e-10-\metric.pdf}
							}
							\hspace{-0.12in}
						}
					\end{minipage}
				}
			}
		}
		\vspace{-0.1in}
		{\includegraphics[scale = 0.4]{\figurepath{\folder}/legend.pdf}}
		\caption{\textbf{Varying $\omega$ on new real datasets under setting of \unstr = $10^6$,  \atnumstr = $6$, \dsstr = $64$, \epstr = $1.0$, \qdimstr = $\qdimlist$. MAEs are shown in log scale.}}
		\label{fig:review_2_comparison_different_approaches_varying_dqv}
	\end{figure*}
}

\def\qdimlist{2, 4}
\def\dqv{0.5}
\def\dqvfile{5}
\def\qn{100}
\foreach \folder/\gran in {all_algorithm_vary_attribute_num}
{
	\begin{figure*}[ht]
		\centering
		\foreach \datasetname/\capdatasetname in {loan/Loan, acs15/Acs}
		{
			\foreach \ds in {64}
			{
				\begin{minipage}[t]{\mpagewidth\linewidth}
					\centering
					\foreach \qdim in \qdimlist
					{
						\subfigure[\textbf{\capdatasetname}, \qdimstr= \qdim] 
						{
							\includegraphics[width=\setwidth\columnwidth, height=\setheight\columnwidth]{\figurepath{\folder}/dn-\datasetname-un-1000000-ds-\ds-fot-2-pf-2-qn-\qn-qd-\qdim-dqv-\dqvfile-e-10-\metric.pdf}
						}
						\hspace{-0.12in}
					}
				\end{minipage}
			}
		}
		\vspace{-0.1in}
		{\includegraphics[scale = 0.4]{\figurepath{\folder}/legend.pdf}}
		\caption{\textbf{Varying $d$ on new real datasets under setting of \unstr = $10^6$,  \dsstr = $64$,  \epstr = $1.0$, \dqvstr = $0.5$, \qdimstr = $\qdimlist$. MAEs are shown in log scale.}}
		\label{fig:review2_comparison_different_approaches_varying_attribute_num}
	\end{figure*}
}

We use two new real datasets to evaluate the performance of HDG.
\begin{itemize}
	\item Loan~\cite{data:loan_new}: It is from the Lending Club and has around 2.2 million records of loan.
	\item Acs~\cite{data:acs}: It is collected by 2015 American Community Survey and has around 1.6 million actual responses.
\end{itemize}

Similar to the previous real datasets Ipums and Bfive, for datasets Loan and Acs, we sample 1 million user records. For evaluation varying different numbers of attributes and domain sizes, we generate multiple versions of these two datasets with the number of attributes ranging from 3 to 10. 

Figures~\ref{fig:review_2_comparison_different_approaches_varying_epsilon}-\ref{fig:review2_comparison_different_approaches_varying_attribute_num} show the results varying $\varepsilon, \omega, d$ when $\lambda = 2, 4$, respectively. We can see that HDG can consistently perform better than the baseline approaches, which further confirms its superiority. We also find that on Loan dataset, HDG achieves higher accuracy than TDG in Figure~\ref{fig:review2_comparison_different_approaches_varying_attribute_num}(a) where $\lambda = 2$, while the situation is reversed in Figure~\ref{fig:review2_comparison_different_approaches_varying_attribute_num}(b) where $\lambda = 4$. It is because using Weighted Update method to estimate the answers of higher dimensional queries may introduce uncertainty on accuracy due to its dependency on the distribution of dataset, as described in Section 4.5. Besides, we have conducted the above experiments under setting of $\lambda= 3, 5, 6$; the results give similar conclusion, and are omitted. We also evaluate the performance of each approach on a new set of synthetic datasets varying the covariance between every two attributes ranging from 0 to 1. The results are shown in Figure~\ref{fig:appendix_synthetic_covariance_varying_epsilon} in Appendix~\ref{appendxi:extra_fig_and_table}, which confirms the superiority of HDG for handling diverse datasets.

\subsection{Maximum Entropy Optimization Method}
\label{appendix:maximum_entropy}

To transform the estimation problem in Section~\ref{subsec:estimation_for_lambda_query} into the Maximum Entropy optimization problem~\cite{DBLP:conf/sigmod/QardajiYL14,DBLP:conf/ccs/ZhangWLHC18}, we first define some necessary notations. Specifically, for a $\lambda$-D range query $q$, we define a set of range queries derived from $q$ as
\begin{equation*}
	Q(q) = \{ { \wedge _t}({a_t},[{l_t},{r_t}] \text{ or } {[{l_t},{r_t}]' }) \mid {a_t} \in {A_q}\},
\end{equation*}
where the interval $[{l_t},{r_t}]'$ is the complement of $[l_t, r_t]$ on the domain of $a_t$. 
Since $A_q$ contains $\lambda$ attributes, there are $2^\lambda$ queries in $Q(q)$. In addition, we define $P_q$ as the set of answers of queries in $Q(q)$. For ease of presentation, we use variable $x$ to denote a query in $Q(q)$.
For any $x \in Q(q)$, we use ${P_q}(x)$ to denote its answer. Similarly, for each 2-D range query $q^{(j, k)}$, the definitions of $Q(q^{(j, k)})$ and $P_{q^{(j, k)}}$ can be obtained. 
In particular, for a $x \in Q(q^{(j, k)})$, ${P_q}(x)$ means $x$'s answer constructed from $P_q$ by summing up the answers of the associated queries in $Q(q)$. 

With the above definitions, we can formulate the problem as the following optimization:
\begin{align*}
	maximize &\quad - \sum\limits_{x \in {Q(q)}} {{P_q}(x) \cdot \log \left( {{P_{q}}(x)} \right)} \\
	subject \,\, to &\quad {\forall _{x \in {Q(q)}}}{P_q}(x) \ge 0 \\
	&\quad {\forall _{q^{(j, k)}}}{\forall _{x \in {Q(q^{(j, k)})}}}{P_{{q^{(j, k)}}}}(x) = {P_q}(x).
\end{align*}
The above optimization problem can be addressed by an off-the-shelf convex optimization tool.

\subsection{Explaining Error Expressions}
We use the following example to explain the expressions of noise and sampling error and non-uniformity error.

\def\folder{visio_figure}
\begin{figure}[htbp]
	\centering
	\includegraphics[scale = 0.7]{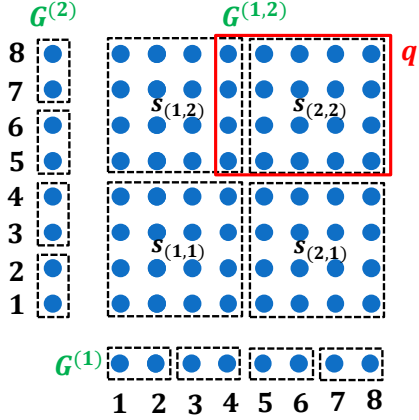}
	\caption{\textbf{Grids regarding $a_1$, $a_2$ and $(a_1, a_2)$.}}
	\label{fig:grid_simple}
\end{figure}

\begin{example}
	\label{example_1}
	
	Assume that the attribute set is $\{a_1, a_2, a_3\}$ with domain size $c = 8$. There are $\binom{3}{2} = 3$ attribute pairs that can be generated from this attribute set in total:
	\begin{equation*}
		(a_1, a_2), \quad (a_1, a_3), \quad (a_2, a_3).
	\end{equation*}
	Figure~\ref{fig:grid_simple} shows the grids regarding $a_1$, $a_2$ and $(a_1, a_2)$ with $g_1 = 4$ and $g_2 = 2$. In Figure~\ref{fig:grid_simple}, the 2-D domain of $(a_1, a_2)$ is partitioned into the $2\times2$  grid $G^{(1, 2)}$ containing $4$ cells $\{s_{(1,1)}, s_{(1,2)}, s_{(2,1)}, s_{(2,2)}\}$. For $a_1$ and $a_2$, their corresponding 1-D grids are $G^{(1)}$ and $G^{(2)}$, each of which also contains $4$ cells.

	Given a range query $q$ which is the red rectangle in Figure~\ref{fig:grid_simple}, it can be answered with uniformity assumption as follows. Since the cell $s_{(2,2)}$ in $G^{(1, 2)}$ is completely included in $q$, its frequency $\eef_{s_{(2,2)}}$ is directly added to the answer $\eef_q$. For the cell $s_{(1,2)}$ that intersects with $q$, the frequencies of four common values between $s_{(1,2)}$ and $q$ should be added to $\eef_q$. With uniformity assumption, the sum of frequencies of these four common values is calculated as $\frac{4}{16} \cdot {\eef_{s_{(1,2)}}} = \frac{1}{4}{f_{{s_{(1,2)}}}}$.
	
	In the estimated answer $\eef_q$, the noise and sampling error comes from the frequency $\eef_{s_{(2,2)}}$ of the cell $s_{(2,2)}$. With the given $\varepsilon$ and derived $\{m_2, n_2\}$ for 2-D grids, we can compute the squared noise and sampling error as $1 \cdot \frac{{4{m_2}{e^\varepsilon }}}{{{n_2}{{({e^\varepsilon } - 1)}^2}}} = \frac{{{m_2}{e^\varepsilon }}}{{{n_2}{{({e^\varepsilon } - 1)}^2}}}$.
	The non-uniformity error is from the estimated sum of frequencies of four common values between $s_{(1,2)}$ and $q$. 
	Assume that the true sum of frequencies of these four common values is $\rf_C$. The squared non-uniformity error equals $\left (\frac{1}{4}{\eef_{{s_{(1,2)}}}} - \rf_C \right)^2$. 
	
	Note that the accurate magnitude of non-uniformity error in a query's estimated answer depends on the true data distribution, which is not available due to LDP guarantee in our problem setting. Therefore, in our guideline, we adopt a simple assumption to measure this error for a general case as described in Section~\ref{subsec:choosing_the_granularity}.
\end{example}

\subsection{Detailed Derivation of Equations}
\label{appendix:detailed_derivation_of_equations}

The Equation~\eqref{equ:wfo:sample:varproof} in Section~\ref{subsec:theoretical_analysis} is
\begin{align}
	& \ep{\left(\eef_v(\sds) - \rf_v\right)^2}\nonumber\\
	=& \ep{\left((\eef_v(\sds) - \rf_v(\sds)) + (\rf_v(\sds) - \rf_v)\right)^2} \nonumber\\
	= & \ep{\left(\eef_v(\sds) - \rf_v(\sds)\right)^2} + \ep{\left(\rf_v(\sds) - \rf_v\right)^2}+\nonumber\\
	& 2\ep{(\eef_v(\sds) - \rf_v(\sds))\cdot(\rf_v(\sds) - \rf_v)} \nonumber \label{equ:wfo:sample:varproof_appendix}
\end{align}
Specifically, the above Equation consists of three parts. The first part is the variance of frequency oracle, i.e.,   
\begin{align}
	& \ep{\left(\eef_v(\sds) - \rf_v(\sds)\right)^2} \nonumber \\
	= &\, m \cdot \frac{p'(1-p') + {\rf}_v(p-p')(1-p-p')}{n(p-p')^2}\nonumber\\
	= &\, m\cdot \frac{p'(1-p')}{n(p-p')^2} + m\cdot \frac{{\rf}_v(p-p')(1-p-p')}{n(p-p')^2}\nonumber
\end{align}

The second part is
\begin{align*}
	&\ep{\left(\rf_v(\sds) - \rf_v\right)^2}\\
	=&\, \ep{\rf^2_v(\sds)} - 2\rf_v\ep{\rf_v(\sds)}+\rf_v^2\\
	=&\, \ep{\rf^2_v(\sds)} - \rf_v^2\\
	=&\, \ep{\left(\frac{m}{n}\sum \ind{v_i=v}\right)^2} - \rf_v^2\\
	=&\left(\frac{k}{n}\right)^2\ep{\left(\sum \ind{v_i=v}\right)^2} - \rf_v^2\\
	=&\, \left(\frac{m}{n}\right)^2\ep{\sum_i \ind{v_i=v}^2 + \sum_{i\neq j} \ind{v_i=v}\cdot \ind{v_j=v}} - \rf_v^2\\
	=&\, \left(\frac{m}{n}\right)^2\left[\frac{n}{m} \rf_v + \left(\frac{n^2}{m^2} - \frac{n}{m}\right) \rf_v\cdot \frac{n\rf_v - 1}{n - 1}\right] - \rf_v^2\\
	=&\frac{k}{n} \rf_v + \left(1 - \frac{k}{n}\right) \rf_v\cdot \frac{n\rf_v - 1}{n - 1} - \rf_v^2\\
	=&\left(\frac{k}{n} - \frac{n - k}{n}\frac{1}{n - 1}\right) \rf_v + \left(1 - \frac{k}{n}\right) \rf_v\cdot \frac{n\rf_v}{n - 1} - \rf_v^2\\
	=&\, \frac{m-1}{n-1} \rf_v (1 - \rf_v).
\end{align*}

The third part is
\begin{align}
	& 2\ep{(\eef_v(\sds) - \rf_v(\sds))\cdot(\rf_v(\sds) - \rf_v)} \nonumber
	\\
	= &\, 2\ep{(\eef_v(\sds) - \rf_v(\sds)) \cdot \rf_v(\sds)}  \nonumber
	\\
	& \hbox{(as $\epinline{\eef_v(D_s)} = \epinline{\rf_v(D_s)}$ and $\rf_v$ is a constant)} \nonumber
	\\
	= &\, 2\ep{\ep{(\eef_v(\sds) - \rf_v(\sds)) \cdot \rf_v(\sds) ~\middle|~ \sds}} \nonumber
	\\
	= &\, 0. \nonumber
\end{align}

\subsection{Extra Figures and Table}
\label{appendxi:extra_fig_and_table}

Figures~\ref{fig:appendix_comparison_different_approaches_varying_epsilon_lambda_6}-\ref{fig:appendix_comparison_different_approaches_varying_user_num_lambda_6} show the results of $\lambda = 6$ varying $\varepsilon, \omega, c, d$ and $n$, respectively.
Figure~\ref{fig:appendix_synthetic_covariance_varying_epsilon} presents the results on  a set of synthetic Normal and Laplace datasets varying the covariance between every two attributes ranging from 0 to 1.
Table~\ref{table:choice_of_granularity} reports the values of $(g_1, g_2)$ used in our experiments under different setting of $\{d, n, \varepsilon\}$.

\def\metric{MNAE}
\def\setwidth{0.47}
\def\setheight{0.36}
\def\mpagewidth{0.48}

\def\setwidth{\fwidth}
\def\setheight{\fheight}

\def\atumlist{6}
\def\dslist{64}
\def\qdimlist{6}
\def\dqv{0.5}
\def\dqvfile{5}
\def\qn{200}

\def\unstr{$n$}
\def\atnumstr{$d$}
\def\dqvstr{$\omega$}
\def\qdimstr{$\lambda$}
\def\qnstr{$|Q|$}
\def\epstr{$\varepsilon$}
\def\dsstr{$c$}

\foreach \folder in {all_algorithm_vary_epsilon}
{
	\begin{figure*}[t]
		\centering
		\foreach \datasetname/\capdatasetname in {ipums2018/Ipums, bfive/Bfive, normal/Normal, laplace/Laplace}
		{
			\foreach \atnum in {6}
			{
				\foreach \ds in {64}
				{
					\centering
					\foreach \qdim in \qdimlist
					{
						\subfigure[\textbf{\capdatasetname}, \qdimstr= \qdim] 
						{
							\includegraphics[width=\setwidth\columnwidth, height=\setheight\columnwidth]{\figurepath{\folder}/dn-\datasetname-un-1000000-an-\atnum-ds-\ds-fot-2-pf-2-qn-\qn-qd-\qdim-dqv-\dqvfile-\metric.pdf}
						}
						\hspace{-0.15in}
					}
				}
			}
		}
		\vspace{-0.1in}
		{\includegraphics[scale = 0.4]{\figurepath{\folder}/legend.pdf}}
		\caption{\textbf{Varying $\varepsilon$ on all datasets under setting of \unstr = $10^6$,  \atnumstr = $6$, \dsstr = $64$, \dqvstr = $0.5$, \qdimstr = $6$. MAEs are shown in log scale.}}
		\label{fig:appendix_comparison_different_approaches_varying_epsilon_lambda_6}
		\vspace{-4pt}
	\end{figure*}
}

\def\metric{MNAE}
\def\setwidth{0.47}
\def\setheight{0.36}
\def\mpagewidth{0.48}

\def\setwidth{\fwidth}
\def\setheight{\fheight}

\def\qdimlist{6}
\def\dqv{0.5}
\def\dqvfile{5}
\def\qn{200}
\foreach \folder/\gran in {all_algorithm_vary_dqv}
{
	\begin{figure*}[htb]
		\centering
		\foreach \datasetname/\capdatasetname in {ipums2018/Ipums, bfive/Bfive, normal/Normal, laplace/Laplace}
		{
			\foreach \atnum in {6}
			{
				\foreach \ds in {64}
				{
					\centering
					\foreach \qdim in \qdimlist
					{
						\subfigure[\textbf{\capdatasetname}, \qdimstr= \qdim] 
						{
							\includegraphics[width=\setwidth\columnwidth, height=\setheight\columnwidth]{\figurepath{\folder}/dn-\datasetname-un-1000000-an-\atnum-ds-\ds-fot-2-pf-2-qn-\qn-qd-\qdim-e-10-\metric.pdf}
						}
						\hspace{-0.15in}
					}
				}
			}
		}
		\vspace{-0.1in}
		{\includegraphics[scale = 0.4]{\figurepath{\folder}/legend.pdf}}
		\caption{\textbf{Varying $\omega$ on all datasets under setting of \unstr = $10^6$,  \atnumstr = $6$, \dsstr = $64$, \epstr = $1.0$, \qdimstr = $6$. MAEs are shown in log scale.}}
		\label{fig:appendix_comparison_different_approaches_varying_dqv_lambda_6}
		\vspace{-4pt}
	\end{figure*}
}

\def\metric{MNAE}
\def\setwidth{0.47}
\def\setheight{0.36}
\def\mpagewidth{0.48}

\def\setwidth{\fwidth}
\def\setheight{\fheight}

\def\qdimlist{6}
\def\atnum{6}
\def\dqv{0.5}
\def\dqvfile{5}
\def\qn{200}
\foreach \folder/\gran in {all_algorithm_vary_domain_size}
{
	\begin{figure*}[t]
		\centering
		\foreach \datasetname/\capdatasetname in {normal/Normal, laplace/Laplace}
		{
			\foreach \ds in {64}
			{
				\centering
				\foreach \qdim in \qdimlist
				{
					\subfigure[\textbf{\capdatasetname}, \qdimstr= \qdim] 
					{
						\includegraphics[width=\setwidth\columnwidth, height=\setheight\columnwidth]{\figurepath{\folder}/dn-\datasetname-un-1000000-an-\atnum-fot-2-pf-2-qn-\qn-dqv-\dqvfile-e-10-qd-\qdim-\metric.pdf}
					}
					\hspace{-0.15in}
				}
			}
		}
		\vspace{-0.1in}
		{\includegraphics[scale = 0.4]{\figurepath{\folder}/legend.pdf}}
		\caption{\textbf{Varying $c$ on synthetic datasets under setting of \unstr = $10^6$,  \atnumstr = $6$,  \epstr = $1.0$, \dqvstr = $0.5$, \qdimstr = $6$. MAEs are shown in log scale.}}
		\label{fig:appendix_comparison_different_approaches_varying_domain_size_lambda_6}
		\vspace{-4pt}
	\end{figure*}
}

\def\metric{MNAE}
\def\setwidth{0.47}
\def\setheight{0.36}
\def\mpagewidth{0.48}

\def\setwidth{\fwidth}
\def\setheight{\fheight}

\def\qdimlist{6}
\def\dqv{0.5}
\def\dqvfile{5}
\def\qn{120}
\foreach \folder/\gran in {all_algorithm_vary_attribute_num}
{
	\begin{figure*}[htbp]
		\centering
		\foreach \datasetname/\capdatasetname in {ipums2018/Ipums, bfive/Bfive, normal/Normal, laplace/Laplace}
		{
			\foreach \ds in {64}
			{
				\centering
				\foreach \qdim in \qdimlist
				{
					\subfigure[\textbf{\capdatasetname}, \qdimstr= \qdim] 
					{
						\includegraphics[width=\setwidth\columnwidth, height=\setheight\columnwidth]{\figurepath{\folder}/dn-\datasetname-un-1000000-ds-\ds-fot-2-pf-2-qn-\qn-qd-\qdim-dqv-\dqvfile-e-10-\metric.pdf}
					}
					\hspace{-0.15in}
				}
			}
		}
		\vspace{-0.1in}
		{\includegraphics[scale = 0.4]{\figurepath{\folder}/legend.pdf}}
		\caption{\textbf{Varying $d$ on all datasets under setting of \unstr = $10^6$,  \dsstr = $64$,  \epstr = $1.0$, \dqvstr = $0.5$, \qdimstr = $6$. MAEs are shown in log scale.}}
		\label{fig:appendix_comparison_different_approaches_varying_attribute_num_lambda_6}
		\vspace{-4pt}
	\end{figure*}
}

\def\metric{MNAE}
\def\setwidth{0.47}
\def\setheight{0.36}
\def\mpagewidth{0.48}

\def\setwidth{\fwidth}
\def\setheight{\fheight}

\def\qdimlist{6}
\def\atnum{6}
\def\dqv{0.5}
\def\dqvfile{5}
\def\qn{200}
\foreach \folder/\gran in {all_algorithm_vary_user_num}
{
	\begin{figure*}[htb]
		\centering
		\foreach \datasetname/\capdatasetname in {normal/Normal, laplace/Laplace}
		{
			\foreach \ds in {64}
			{
				\centering
				\foreach \qdim in \qdimlist
				{
					\subfigure[\textbf{\capdatasetname}, \qdimstr= \qdim] 
					{
						\includegraphics[width=\setwidth\columnwidth, height=\setheight\columnwidth]{\figurepath{\folder}/dn-\datasetname-an-\atnum-ds-\ds-fot-2-pf-2-qn-\qn-dqv-\dqvfile-e-10-qd-\qdim-\metric.pdf}
					}
					\hspace{-0.15in}
				}
			}
		}
		\vspace{-0.1in}
		{\includegraphics[scale = 0.4]{\figurepath{\folder}/legend.pdf}}
		\caption{\textbf{Varying $n$ on synthetic datasets under setting of \atnumstr = $6$, \dsstr = $64$, \epstr = $1.0$, \dqvstr = $0.5$, \qdimstr = $6$. MAEs are shown in log scale.}}
		\label{fig:appendix_comparison_different_approaches_varying_user_num_lambda_6}
		\vspace{-9pt}
	\end{figure*}
}


\def\metric{MNAE}
\def\setwidth{0.47}
\def\setheight{0.36}

\def\setwidth{0.23}
\def\setheight{0.14}
\def\mpagewidth{1.0}

\def\atumlist{6}
\def\dslist{64}
\def\qdimlist{2, 4, 6}
\def\dqv{0.5}
\def\dqvfile{5}
\def\qn{200}

\foreach \folder in {all_algorithm_vary_epsilon}
{
	\begin{figure*}[t]
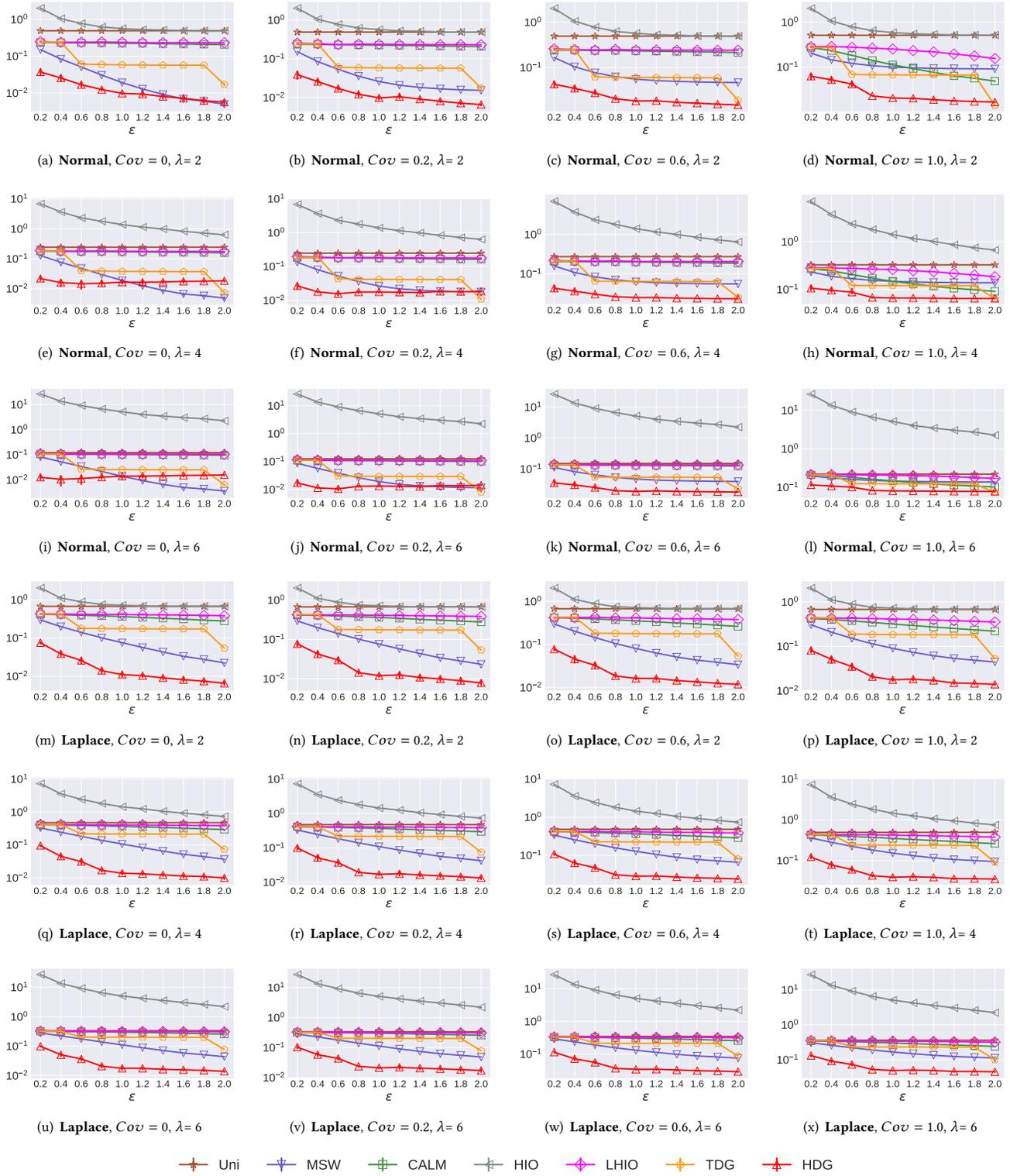

		\centering
		\foreach \qdim in \qdimlist
		{
			\foreach \atnum in {6}
			{
				\foreach \ds in {64}
				{
					\begin{minipage}[t]{\mpagewidth\linewidth}
						\centering
						\foreach \datasetname/\capdatasetname in {normal0/0, normal2/0.2, normal6/0.6, normal10/1.0}
						{
							\subfigure[\textbf{Normal}, $Cov =$ \capdatasetname, \qdimstr= \qdim] 
							{
								\includegraphics[width=\setwidth\columnwidth, height=\setheight\columnwidth]{\figurepath{\folder}/dn-\datasetname-un-1000000-an-\atnum-ds-\ds-fot-2-pf-2-qn-\qn-qd-\qdim-dqv-\dqvfile-\metric.pdf}
							}
						}
					\end{minipage}
				}
			}
		}
		\foreach \qdim in \qdimlist
		{
			\foreach \atnum in {6}
			{
				\foreach \ds in {64}
				{
					\begin{minipage}[t]{\mpagewidth\linewidth}
						\centering
						\foreach \datasetname/\capdatasetname in {laplace0/0, laplace2/0.2,  laplace6/0.6, laplace10/1.0}
						{
							\subfigure[\textbf{Laplace}, $Cov =$ \capdatasetname, \qdimstr= \qdim] 
							{
								\includegraphics[width=\setwidth\columnwidth, height=\setheight\columnwidth]{\figurepath{\folder}/dn-\datasetname-un-1000000-an-\atnum-ds-\ds-fot-2-pf-2-qn-\qn-qd-\qdim-dqv-\dqvfile-\metric.pdf}
							}
						}
					\end{minipage}
				}
			}
		}
		\vspace{-0.1in}
		{\includegraphics[scale = 0.4]{\figurepath{\folder}/legend.pdf}}
		\caption{\textbf{Normal and Laplace datasets with different covariance between every two attributes: comparison varying $\varepsilon$ under setting of \unstr = $10^6$,  \atnumstr = $6$, \dsstr = $64$, \dqvstr = $0.5$, \qdimstr = $2, 4, 6$. MAEs are shown in log scale.}}
		\label{fig:appendix_synthetic_covariance_varying_epsilon}
	\end{figure*}
}

\begin{table*}[ht]
	\centering
	\begin{tabular}{|c|c|c|c|c|c|c|c|c|c|c|}
		\hline
		\diagbox{$d, lg(n)$}{$(g_1, g_2)$}{$\varepsilon$} & 0.2 & 0.4 & 0.6 & 0.8 & 1.0 & 1.2 & 1.4 & 1.6 & 1.8 & 2.0 \\
		\hline
		3 , $6$ & 8, 2 & 16, 4 & 32, 4 & 32, 4 & 32, 4 & 32, 4 & 32, 8 & 64, 8 & 64, 8 & 64, 8 \\ \hline
		4 , $6$ & 8, 2 & 16, 2 & 16, 4 & 32, 4 & 32, 4 & 32, 4 & 32, 4 & 32, 4 & 32, 8 & 64, 8 \\ \hline
		5 , $6$ & 8, 2 & 16, 2 & 16, 4 & 16, 4 & 32, 4 & 32, 4 & 32, 4 & 32, 4 & 32, 4 & 32, 8 \\ \hline
		6 , $6$ & 8, 2 & 16, 2 & 16, 2 & 16, 4 & 16, 4 & 32, 4 & 32, 4 & 32, 4 & 32, 4 & 32, 4 \\ \hline
		7 , $6$ & 8, 2 & 8, 2 & 16, 2 & 16, 4 & 16, 4 & 32, 4 & 32, 4 & 32, 4 & 32, 4 & 32, 4 \\ \hline
		8 , $6$ & 8, 2 & 8, 2 & 16, 2 & 16, 2 & 16, 4 & 16, 4 & 32, 4 & 32, 4 & 32, 4 & 32, 4 \\ \hline
		9 , $6$ & 8, 2 & 8, 2 & 16, 2 & 16, 2 & 16, 4 & 16, 4 & 16, 4 & 32, 4 & 32, 4 & 32, 4 \\ \hline
		10 , $6$ & 4, 2 & 8, 2 & 8, 2 & 16, 2 & 16, 2 & 16, 4 & 16, 4 & 32, 4 & 32, 4 & 32, 4 \\ \hline
		6 , 5.0 & 4, 2 & 4, 2 & 8, 2 & 8, 2 & 8, 2 & 16, 2 & 16, 2 & 16, 2 & 16, 2 & 16, 4 \\ \hline
		6 , 5.2 & 4, 2 & 8, 2 & 8, 2 & 8, 2 & 16, 2 & 16, 2 & 16, 2 & 16, 4 & 16, 4 & 16, 4 \\ \hline
		6 , 5.4 & 4, 2 & 8, 2 & 8, 2 & 16, 2 & 16, 2 & 16, 2 & 16, 4 & 16, 4 & 16, 4 & 32, 4 \\ \hline
		6 , 5.6 & 4, 2 & 8, 2 & 8, 2 & 16, 2 & 16, 2 & 16, 4 & 16, 4 & 32, 4 & 32, 4 & 32, 4 \\ \hline
		6 , 5.8 & 8, 2 & 8, 2 & 16, 2 & 16, 2 & 16, 4 & 16, 4 & 32, 4 & 32, 4 & 32, 4 & 32, 4 \\ \hline
		6 , 6.0 & 8, 2 & 16, 2 & 16, 2 & 16, 4 & 16, 4 & 32, 4 & 32, 4 & 32, 4 & 32, 4 & 32, 4 \\ \hline
		6 , 6.2 & 8, 2 & 16, 2 & 16, 4 & 16, 4 & 32, 4 & 32, 4 & 32, 4 & 32, 4 & 32, 4 & 32, 8 \\ \hline
		6 , 6.4 & 8, 2 & 16, 2 & 16, 4 & 32, 4 & 32, 4 & 32, 4 & 32, 4 & 32, 8 & 64, 8 & 64, 8 \\ \hline
		6 , 6.6 & 16, 2 & 16, 4 & 32, 4 & 32, 4 & 32, 4 & 32, 4 & 32, 8 & 64, 8 & 64, 8 & 64, 8 \\ \hline
		6 , 6.8 & 16, 2 & 16, 4 & 32, 4 & 32, 4 & 32, 4 & 64, 8 & 64, 8 & 64, 8 & 64, 8 & 64, 8 \\ \hline
		6 , 7.0 & 16, 2 & 32, 4 & 32, 4 & 32, 4 & 64, 8 & 64, 8 & 64, 8 & 64, 8 & 64, 8 & 64, 8 \\ \hline
		
	\end{tabular}
	\caption{\textbf{HDG: The recommended granularity settings with fixed $\alpha_1 = 0.7$ and $\alpha_2 = 0.03$. Each cell is a tuple of $(g_1, g_2)$. Each row represents the values for the same $d$ and $lg(n)$ setting.}}
	\label{table:choice_of_granularity}
\end{table*}

\end{document}